\documentclass[runningheads]{llncs}
\usepackage{graphicx,hyperref,xspace,tikz} 
\usepackage{amsmath,amssymb,amsfonts}
\usepackage[todonotes={textsize=footnotesize}]{changes}
\usepackage{subcaption}

\usetikzlibrary{matrix}
\usetikzlibrary{external}
\definechangesauthor[color=red]{LC}
\definechangesauthor[color=blue]{GM}

\definecolor{darkpastelgreen}{rgb}{0.01, 0.5, 0.24}

\long\def\ignora#1{}

\newcommand{\lmax}{\ell_{\max}}
\newcommand{\Maa}{A_{n}}
\newcommand{\bDD}{b_{2D}}
\newcommand{\gammaDD}{\gamma_{2D}}
\newcommand{\GammaDD}{\Gamma_{2D}}
\newcommand{\deltaDD}{\delta_{2D}}
\newcommand{\BTDD}{{\textit{2D-BT}}\xspace}

\title{The landscape of compressibility measures for two-dimensional data\thanks{Partially supported by the INdAM-GNCS Project CUP E53C23001670001, by the spoke ``FutureHPC \& BigData'' of the ICSC --- Centro Nazionale di Ricerca in High-Performance Computing, Big Data and Quantum Computing, funded by European Union NextGeneration EU programme, and by MIUR PRIN Project 2017WR7SHH. This paper improves and extends some results appearing in~\cite{CarfagnaM23}, available at \url{https://doi.org/10.1007/978-3-031-43980-3_9}}}

\author{Lorenzo Carfagna\orcidID{0009-0005-9591-057X} \and \\
Giovanni Manzini\orcidID{0000-0002-5047-0196}}
\authorrunning{L. Carfagna, G. Manzini}

\institute{University of Pisa\\
\email{lorenzo.carfagna@phd.unipi.it\;
giovanni.manzini@unipi.it}}

\begin{document}

\maketitle

\begin{abstract}
In this paper we extend to two-dimensional data two recently introduced one-dimensional compressibility measures: the $\gamma$ measure defined in terms of the smallest {string attractor}, and the $\delta$ measure defined in terms of the number of distinct substrings of the input string. Concretely, we introduce the two-dimensional measures $\gammaDD$ and $\deltaDD$, as natural generalizations of $\gamma$ and~$\delta$, and we initiate the study of their properties. Among other things, we prove that $\deltaDD$ is monotone and can be computed in {linear time}, and we show that, although it is still true that $\deltaDD \leq \gammaDD$, the gap between the two measures can be $\Omega(\sqrt{n})$ and therefore asymptotically larger than the gap between $\gamma$ and~$\delta$. To complete the scenario of two-dimensional compressibility measures, we introduce the measure $\bDD$ which generalizes to two dimensions the notion of optimal parsing. We prove that, somewhat surprisingly, the relationship between $\bDD$ and $\gammaDD$ is significantly different than in the one-dimensional case. As an application of our results we provide the first analysis of the space usage of the two-dimensional block tree introduced in [Brisaboa {\em et al.}, Two-dimensional block trees, {\em The computer Journal}, 2024]. Our analysis shows that the space usage can be bounded in terms of both $\gammaDD$ and $\deltaDD$. Finally, using insights from our analysis, we design the first linear time and space algorithm for constructing the two-dimensional block tree for arbitrary matrices.

\end{abstract}

\section{Introduction}

Since the recent introduction of the notion of string attractor~\cite{10.1145/3188745.3188814} different measures of string repetitiveness have been proposed or revisited~\cite{titt22,Navacmcs20.2}. It has been shown that such measures are more appropriate than the classical statistical entropy for measuring the compressibility of highly repetitive strings and their study has lead to some deep theoretical results. For example they have been used to devise efficient compressed indices for highly repetitive string collections~\cite{Navacmcs20.3} an important setting which is hard for traditional entropy-based compressed indices. 

In this paper we generalize the notion of attractor to two dimensional data, i.e. (square) matrices of symbols, and we initiate the study of the properties of the measure $\gammaDD(M)$ defined as the size of the smallest attractor for the matrix~$M$ (Definition~\ref{def:gamma}). As in the one-dimensional case, we introduce also the measure $\deltaDD(M)$ defined in terms of the number of distinct square submatrices (Definition~\ref{def:delta}) and we study the relationship between $\gammaDD$ and $\deltaDD$. In Section~\ref{sec:2Dattractor} we prove that some properties that hold for strings are still valid in the two-dimensional case: for example computing $\gammaDD$ is NP complete while $\deltaDD$ can be computed in linear time, and for every matrix $M$ it is $\deltaDD(M) \leq \gammaDD(M)$. However, the gap between the two measures is larger than in one-dimensional case since there are families of $n\times n$ matrices with $\deltaDD = O(1)$ and $\gammaDD = \Omega(\sqrt{n})$, whereas for strings it is always $\gamma = O(\delta\log \frac{n}{\delta})$. We also prove that for three dimensional structures the gap is $\Omega(n^{2/3})$ and we conjecture that the gap increases with the number of dimensions.

In Section~\ref{sec:bidi} we introduce another measure for two-dimensional data, called $\bDD$, which generalizes the notion of minimal bidirectional macro scheme defined for strings in~\cite{storer82}. We prove that, while in one dimension the minimal attractor is always smaller than the minimal bidirectional macro scheme, this is not always true in two dimensions: indeed there are families of $n \times n$ matrices for which the ratio $\gammaDD/\bDD$ is $\Omega(\sqrt{n})$.

Having introduced the measures $\gammaDD$, $\deltaDD$, and $\bDD$ and some of their basic properties, we use them for the analysis of a recently introduced compressed data structure: the {\em two-dimensional block tree}~\cite{2dbt}. In Section~\ref{sec:2DBT} we provide the first analysis of the space usage of such data structure as a function of the ``information content'' of the input. 
In particular we show that the space used by a two-dimensional block tree for an $n\times n$ matrix $M$ with delta measure $\deltaDD$ is bounded by $O((\deltaDD + \sqrt{n\deltaDD})\log ({n}/{\sqrt{\deltaDD}}))$ and that this space is optimal within a multiplicative factor $O(\log n)$. We also show how to build a space efficient two-dimensional block tree given a two-dimensional attractor, and we use this result to prove a new relationship between the measures $\gammaDD$ and $\bDD$. 
Finally, in Section~\ref{sec:2DBT-con} we provide the first optimal linear time and space algorithm to build the two-dimensional block tree. 

We believe that our theoretical results, combined with the good practical performance reported in~\cite{2dbt}, show the attractiveness of the two-dimensional block tree and its suitability to become a general purpose tool to efficiently represent compressible two dimensional structures. 

\section{Notation and background}\label{sec:notation}

In this paper we consider one-dimensional strings and two-dimensional matrices over an integer alphabet $\Sigma$ with $|\Sigma|=\sigma$.  Given the string $S\in\Sigma^n$ we denote its symbols with $S[i]$ for $i=1,\ldots,n$, and we write $S[i..j]$ to denote the substring $S[i]S[i+1]\cdots S[j]$. Given the matrix $M\in \Sigma^{m\times n}$ we denote its symbols with {$M[i][j]$} for $i=1,\ldots,m$, $j=1,\ldots,n$. A submatrix of $M$ with topmost left cell $M[i][j]$ is said to start at position $(i,j)$ of $M$. An $a \times b$ submatrix of $M$ starting at position $(i,j)$ is written as $M[i..i+a-1][j..j+b-1]$, meaning that it includes any cell with row index in the range $[i,i+a-1]$ and column index in $[j,j+b-1]$. We assume that $\Sigma$ is effective, that is every symbol in $\Sigma$ appears in $M$ (or in~$S$). Therefore, when the input is an $n\times m$ matrix, we will assume $\sigma\leq nm$.

For the rest of the paper, the RAM model of computation is assumed, with word size $w=\Theta(\log n)$ bits. Space is measured in words so when $O(x)$ space is indicated, the actual space occupancy in bits is $O(x \log n)$.

We now recall the definition of the $\gamma$ and $\delta$ measures for strings respectively introduced in~\cite{10.1145/3188745.3188814} and \cite{10.1145/3426473,10.1007/s00453-012-9618-6}. 

\begin{definition}\label{def:gamma1}
An attractor $\Gamma$ for a string $S[1..n]$ is a set of positions $\Gamma \subseteq \{1,...,n\}$ such that any substring of $S$ has an occurrence (i.e. a copy) crossing (i.e. including) a position $p \in \Gamma$. The measure $\gamma(S)$ is defined as the cardinality of a smallest attractor for $S$. 
\end{definition}

\begin{definition} \label{def:delta1}
Given $S[1..n]$, let $d_{k}(S)$ be the number of distinct $k$ substring of $S$, then
\begin{equation*}
    \delta(S)= \max \{d_{k}(S)/{k}\colon k \in [1, n]\}
\end{equation*}
\end{definition}

It is known~\cite{10.1007/s00453-012-9618-6} that for any string $S$ it is $\delta(S) \leq \gamma(S)$, and while computing $\gamma(S)$ is NP complete, $\delta(S)$ can be computed in linear time. In addition, $\delta$ is {\em monotonic} in the sense that if $S'$ is a substring of $S$ then it is $\delta(S') \leq \delta(S)$.  

A \emph{bidirectional macro scheme}~\cite{storer82} consists of a partition of the input string $S[1..n]$ into a sequence of non overlapping phrases $S[1..i_1]S[i_1+1..i_2]\cdots S[i_m+1..n]$ such that each phrase is either an explicit symbol (i.e. it is has length 1) or is a copy of another substring. The only restriction on the copy operations is that it must be possible to eventually retrieve the value of any entry in $S$. Formally, given a macro scheme we define a function $f\colon \{1,\ldots,n\} \to \{0,\ldots,n\}$ such that: $f(i)=0$ if $S[i]$ is an explicit symbol, and $f(i) = i+j$, if $S[i]$ belongs to a phrase $S[u..v]$ which is copied from $S[u+j..v+j]$ (note that $j$ can be either positive and negative). The macro scheme is \emph{valid} only if for $i=1,\ldots,n$ there exists $k\geq 1$ such that $f^k(i)=0$, where $f^k$ denotes the function $f$ iterated $k$ times. 

\begin{definition}\label{def:macros1D}
    For any string $S$, we write $b(S)$, or simply $b$ when $S$ is clear from the context, to denote the minimum number of phrases in a valid bidirectional macro scheme for~$S$. 
\end{definition}

The value $b(S)$ is a natural measure of complexity of the string $S$. Indeed, the value $b(S)$ is a lower bound to the number of phrases of many parsing schemes used for data compression; at the same time $b(S)$ is a reachable measure, in the sense that one can encode $S$ with $O(b(S))$ words. On the negative side we have that $b(S)$ is not monotone, and determining $b(S)$ given $S$ is NP-hard~\cite{Gal82}. Finally, in~\cite{Navacmcs20.3} it is shown that for any string $S$ it is $\gamma(S) \leq b(S)$.

The measures $\gamma(S)$, $\delta(S)$ and $b(S)$ have been introduced as alternative to the {\em statistical $k$-th order entropy} $H_k(S)$ which measures how well each symbol is predicted by its length-$k$ context, i.e. the $k$ symbols preceding it in $S$. This entropy is also called {\em empirical} since it is derived from probabilities observed on the input string $S$ without assuming that $S$ is generated by a random source.  Although the statistical entropy remains a cornerstone in many fields, including the study of compressed data structures~\cite{cds}, it has been observed in~\cite[Sect.~3.1]{Navacmcs20.3} that such measure fails to fully capture the regularities of inputs consisting of a repeated pattern with small variants, such as collections of genomes of the same species or software repositories storing all the successive versions of every file. In~\cite{Navacmcs20.3} and~\cite{Navacmcs20.2} it is given theoretical and practical evidence that in those settings $\gamma(S)$, $\delta(S)$ and $b(S)$ provide a better measure of the ``information content'' of the input.

In the domain of two-dimensional structures, the interest in alternative compressibility measures is even greater because there is not a widely accepted notion of empirical entropy, mainly due to the absence of a clear definition of the 'context' of a symbol. To our knowledge, the only proposal of a two-dimensional compressibility measure analogous to the entropy is the measure $H^*(I)$ proposed in~\cite{Lempel_Ziv_1986} for an {\em infinite} matrix~$I$. Informally, $H^*(I)$ is defined partitioning the $N\times N$ upper left corner of $I$ into $q\times q$ submatrices, with $q=2^j$, and measuring the resulting statistical entropy $H^*_q(I_N)$ defined as follows: If the $i$-th $q\times q$ submatrix occurs $W_i$ times out of $W=\sum_i W_i$ its contribution to $H^*_q(I_N)$ is $-(q^2 \log|\Sigma|)^{-1} (W_i/W)\log(W_i/W)$. From $H^*_q(I_N)$ the measure $H^*(I)$ is defined taking the limit for $N\to\infty$ and $q\to\infty$. Although~\cite{Lempel_Ziv_1986} shows that $H^*(I)$ has intriguing theoretical properties, it is not obvious how such measure can be used to estimate the compressibility of a {\em finite} matrix.

After the publication of the preliminary version of this paper~\cite{CarfagnaM23}, Romana et al.~\cite{Romana_Sciortino_Urbina_2024}, proposed and analyzed new 2D-measures that generalise and extend those presented in this paper. We postpone the discussion of Romana et al.'s results to the Section~\ref{sec:concluding} dedicated to the concluding remarks.

\section{Attractors for two-dimensional structures}\label{sec:2Dattractor}

In this section we generalize the notion of attractor to matrices, and we introduce two new repetitiveness measures for square matrices, called $\gamma_{2D}$ and $\delta_{2D}$, as the generalisations of the $\gamma$ and $\delta$ measures for strings. 

\begin{definition}\label{def:gamma}
An attractor $\Gamma_{2D}$ for a square matrix $M\in \Sigma^{n\times n}$ is a set of positions $\Gamma_{2D} \subseteq \{1,...,n\}\times \{1,...,n\}$ such that any square submatrix of $M$ has an occurrence crossing (i.e. including) a position $p = (i,j) \in \Gamma_{2D}$. The measure $\gamma_{2D}(M)$ is defined as the cardinality of a smallest attractor for $M$. 
\end{definition}
We say that a position $p=(i,j)\in \Gamma_{2D}(M)$ \textit{covers} a submatrix $I$ of $M$ if there exists an occurrence of $I$ which crosses $p$, and that a set of positions \textit{covers} $I$ if it includes a position $p$ which \textit{covers} $I$; when clear from the context, the parameter $M$ is omitted from $\Gamma_{2D}(M)$ expression.

As a first result we show that, not surprisingly, the problem of finding the size of a smallest attractor is \textit{NP-complete} also in two dimensions. The {NP-completeness} proof is done considering the decision problem ``is there an attractor of size $k$ for the given input?''.

\begin{lemma}\label{lemma:reduction}
Given a string $S \in \Sigma^n$, let $R^S \in \Sigma^{n\times n}$ be the square matrix where each row is equal to the string $S$. Then there exists an (1-dim) attractor for $S$ of size $k$ if and only if there exists a (2-dim) attractor of size $k$ for $R^S$.
\end{lemma}

\begin{proof}
Given $S$ and the corresponding $R^S$, the following observations hold: $1)$ any submatrix of $R^S$ has an occurrence starting at the same column but on the first row of $R^S$; $2)$ any two $\ell \times \ell$ submatrices of $R^S$ are equal if and only if the two respective substrings of $S$ composing their rows are equal, formally: $R^S[i..i+\ell-1][j..j+\ell-1]=R^S[i'..i'+\ell-1][j'..j'+\ell-1]$ if and only if $S[j..j+\ell-1]=S[j'..j'+\ell-1]$.
From $1)$ and $2)$ the lemma follows: given a string attractor $\Gamma(S)$ for $S$ of size $k$, the set $\GammaDD = \{(1,j)\colon j \in \Gamma(S)\}$ of size $k$ is a two dimensional attractor for $R^S$ and, vice versa, a string attractor $\Gamma$ for $S$ could be obtained from a matrix attractor $\Gamma_{2D}(R^S)$ for $R^S$ projecting each couple by column index, formally, $\Gamma=\{j\colon (i,j) \in \Gamma_{2D}(R^S)\}$ is a one dimensional attractor for $S$. Note that if $\Gamma_{2D}(R^S)$ is a smallest attractor it does not include two positions on the same column, because, any distinct submatrix crossing one position has an occurrence (starting in the same column but at a different row) which crosses the other, hence in this case the projection does not generate any column index collision and $|\Gamma|=|\Gamma_{2D}(R^S)|=k$, otherwise, in case of collision, $\Gamma$ could be completed with any $k-|\Gamma|$ positions not in $\Gamma$ to reach size $k=|\Gamma_{2D}(R^S)|$. \qed  
\end{proof}

As an immediate consequence of the above lemma we have the following result. 

\begin{theorem}
    Computing $\gammaDD$ is NP complete.
\end{theorem}

It is easy to see that $\gamma_{2D} \geq \sigma$ (the alphabet size) and $\gamma_{2D}$ is insensitive to transpositions but, as for strings~\cite{10.1016/j.tcs.2020.11.006}, $\gamma_{2D}$ is not monotone. We show this by providing a family of matrices, built using the counterexample in~\cite{10.1016/j.tcs.2020.11.006} to disprove the monotonicity of $\gamma$, containing a submatrix with smaller $\gamma_{2D}$.

\begin{lemma}
$\gamma_{2D}$ is not monotone.
\end{lemma}

\begin{proof}
Let $w$ be the string $a\underline{b}b\underline{b}a^n\underline{a}b$ {with $n>0$,} having $\gamma(w)=3$ minimal for the subset of positions $\Gamma(w) = \{2,4,n+5\}$ underlined in $w$. The string $w\cdot b = abb\underline{b}a^n\underline{a}bb $ obtained concatenating the letter $b$ to $w$ has a smaller compressibility measure $\gamma(w\cdot b)=2$ corresponding to $\Gamma(w\cdot b) = \{4,n+5\}$ \cite{10.1016/j.tcs.2020.11.006}, as the prefix $w[1,3]=abb$ occurring as a suffix of $w\cdot b$ is already covered by position $n+5$ in $\Gamma(w\cdot b)$. 
Consider $R^{w\cdot b}$ of size $(n+7)\times (n+7)$, from Lemma~\ref{lemma:reduction} follows that $\gammaDD(R^{w\cdot b})=\gamma(w\cdot b)=2$, but the submatrix $R^{w\cdot b}[1..n+6][1..n+6]$ equal to $R^w$ has a greater $\gammaDD(R^w)=\gamma(w)=3$. \qed
\end{proof}

\subsection{The measure $\deltaDD$}\label{sec:delta2D}

The measure $\delta(S)$ for a string $S$, formally defined in~\cite{10.1145/3426473} and previously introduced in~\cite{10.1007/s00453-012-9618-6} to approximate the output size of the Lempel–Ziv parsing, is the maximum over $k\in [1,|S|]$ of the expression $d_k(S)/k$ where $d_k(S)$ is the number of distinct substrings of length $k$ in $S$. We now show how to generalize this measure to two dimensions, by introducing the measure  $\delta_{2D}$ which is defined in a similar way, considering $ k \times k$ submatrices instead of length-$k$ substrings.

\begin{definition} \label{def:delta}
Given $M\in\Sigma^{n\times n}$, let $d_{k\times k}(M)$ be the number of distinct $k\times k$ submatrices of $M$, then
\begin{equation*}
    \delta_{2D}(M)= \max \{d_{k\times k}(M)/{k^2}\colon k \in [1 , n]\}
\end{equation*}
\end{definition}

The measure $\delta_{2D}$ preserves some good properties of $\delta$: $\delta_{2D}$ is invariant through transpositions and decreases or grows by at most $1$ after a single cell edit since any $d_{k\times k}$ of the updated matrix could differ at most by $k^2$ from the initial one. $\delta_{2D}$ is monotone: to see this observe that given a submatrix $M'$ of $M$ having size $\ell \times \ell$ with $\ell\leq n$ any submatrix of $M'$ appears somewhere in $M$, hence $d_{k\times k}(M')\leq d_{k\times k}(M)$ for any $k\in[1,\ell]\subseteq [1,n]$. 

The next lemma shows that, as in the one-dimensional setting, $\delta_{2D}$ is upper bounded by~$\gamma_{2D}$.

\begin{lemma} \label{lemma:delta<gamma}
$\delta_{2D}(M)\leq \gamma_{2D}(M)$ for any matrix $M\in \Sigma^{n\times n}$
\end{lemma}
\begin{proof}
Let $\GammaDD$ be a least size attractor for $M$ i.e. $|\GammaDD|=\gamma_{2D}$. For any $k \in [1,n]$ an attractor position $p \in \GammaDD$ is included in at most $k^2$ distinct {$k \times k$} submatrices, then we need at least $d_{k\times k}(M)/k^2$ distinct positions in $\GammaDD$ to cover all $k\times k$ submatrices of $M$, formally, $|\GammaDD| \geq d_{k\times k}(M)/k^2$ holds for any $k\in[1,n]$ in particular for $k^*\in[1,n]$ such that $\delta_{2D}= d_{k^*\times k^*}(M)/{(k^*)}^2$. \qed
\end{proof}

One of the main reasons for introducing~$\delta$ was that it can be computed efficiently: \cite{10.1145/3426473} describes a linear algorithm to compute $\delta(S)$ with a single visit of the Suffix tree of $S$. We now show that an efficient algorithm for computing $\delta_{2D}$ can be derived in a similar way using the \textit{Isuffix tree} introduced in~\cite{icalp99} which can be built in $O(n^2)$ time (see~\cite[Theorem~1]{algorithmica/KimNSP11}), which is linear in the size of the input. A somewhat simpler algorithm can be obtained using the \textit{Lsuffix tree} of~\cite{siamcomp/Giancarlo95} but its construction takes $O(n^2\log n)$ time.

\begin{figure}[t]
    \centering
    \includegraphics[]{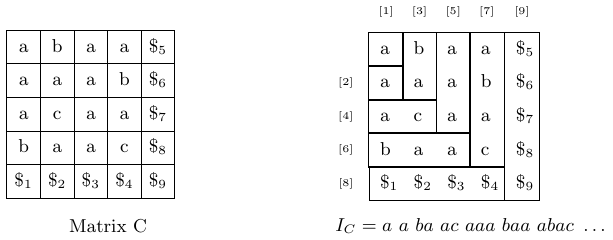}
    \caption{A square matrix $C$ on the left, and its Istring $I_C$ on the right (last two Icharacters are omitted)} 
    \label{fig:M1}
\end{figure}

The \textit{Isuffix tree} $IST(A)$ of a matrix $A\in\Sigma^{n\times m}$ generalises the Suffix Tree to matrices: $IST(A)$ is a compacted trie representing all square submatrices of $A$.
The \textrm{Isuffix trees} adopts a linear representation of a square matrix $C\in\Sigma^{q\times q}$: let $I \Sigma = \bigcup_{i=1}^{\infty}\Sigma^{i}$, each string in $I \Sigma$ is considered as an atomic \textit{Icharacter}, the unique \textit{Istring} associated to matrix $C$ is $I_C\in {I\Sigma}^{2q-1}$ where $I_C[2i+1]$ with $i\in [0,q)$ is the ${(i+1)}^{th}$ \textit{column-type Icharacter} $C[1..i+1][i+1]$ and $I_C[2i]$ with $i\in [1,q)$ is the $i^{th}$ \textit{row-type Icharacter} $C[i+1][1..i]$. See Figure~\ref{fig:M1} for an example. 
The $k^{th}$ \textit{Iprefix} of $C$ is defined as the concatenation of the first $k$ \textrm{Icharacters} $I_C[1]\cdot I_C[2] \cdot \ldots \cdot I_C[k]=I_C[1,k]$ of $I_C$. Note that an \textrm{Iprefix} ending in an odd position $k$ is the \textrm{Istring} of the $\ell \times \ell$ square submatrix with $\ell=\lceil k/2 \rceil$ starting at $C$'s top-left corner, that is, $C[1..\ell][1..\ell]$. For the example in Figure~\ref{fig:M1}, the $3^{rd}$ Iprefix of $C$ is $a\; a\; ba$ which corresponds to the submatrix $C[1..2][1..2]$.

\begin{figure}[t]
    \centering
    \vspace{3pt}
    \includegraphics[]{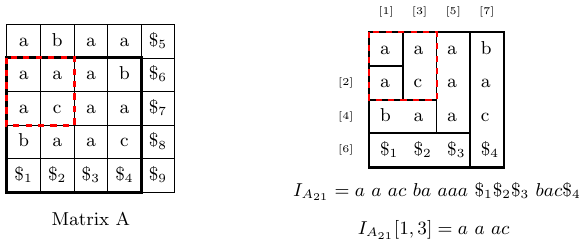}
    \caption{The submatrix $A[2..5][1..4]=A_{21}$ with solid black border on the left and its Istring $I_{A_{21}}$ on the right. The Istring of the submatrix $A[2..3][1..2]$ (in red) is the third Iprefix of $I_{A_{21}}$.} 
    \label{fig:M2}
\end{figure}

Given $A\in\Sigma^{n\times n}$, for $1 \leq i,j \leq n$, the \textit{Isuffix} $I_{A_{ij}}$ of $A$ is defined as the \textrm{Istring} of the largest square submatrix $A_{ij}$ of $A$ with upper left corner at position $(i,j)$. From the above definitions is clear that the \textrm{Istring} of any square submatrix of $A$, is an \textrm{Iprefix} (ending in a odd position) of some \textrm{Isuffix} $I_{A_{ij}}$. To ensure that no \textrm{Isuffix} $I_{A_{ij}}$ is \textrm{Iprefixed} by another \textrm{Isuffix}, $A$ is completed with an additional bottom row and right column containing $2n+1$ distinct new symbols $\$_1, \ldots \$_{2n+1}$. For simplicity in the following we refer as $A$ the input matrix already enlarged with $\$_i$ symbols. See Figure~\ref{fig:M2} for an example.

The \textrm{Isuffix tree} $IST(A)$ is a compacted trie over the alphabet $I\Sigma$ representing all the $n^2$ distinct \textrm{Isuffixes} $I_{A_{ij}}$ of $A$ with, among others, the following properties~\cite{algorithmica/KimNSP11}: 1) each edge is labeled with a non empty \textrm{Isubstring} $I_{A_{ij}}[\ell_1,\ell_2]$ of an \textrm{Isuffix} $I_{A_{ij}}$, that label is represented in constant space as the quadruple $\langle i,j,\ell_1,\ell_2 \rangle$, the \textrm{Isubstrings} on any two sibling edges start with different \textrm{Icharacters}; 2) each internal node has at least two children and there are exactly $n^2$ leaves representing all the \textrm{Isuffixes} of $A$: let $L(u)$ be the \textrm{Istring} obtained concatenating the \textrm{Isubstrings} on the path from the root to a node $u$, for any leaf $l_{ij}$, the \textrm{Istring} $L(l_{ij})$ is equal to the linear representation $I_{A_{ij}}$ of the unique suffix $A_{ij}$; 3) The \textrm{Isuffix tree} satisfies the \textit{common prefix constraint:} square submatrices of $A$ with a common \textrm{Iprefix} share the same initial path in the tree; 4) The \textrm{Isuffix tree} satisfies the \textit{completeness constraint} since all square submatrices of $A$ are represented in $IST(A)$ as an \textrm{Iprefix} of some \textrm{Isuffix} of $A$.

\begin{theorem}
$\delta_{2D}$ can be calculated in optimal time and space $O(n^2)$
\end{theorem}

\begin{proof}
Our algorithm is a generalization of the ideas used in~\cite{10.1145/3426473} to compute the measure $\delta$ in linear time using a suffix tree. Given $A\in\Sigma^{n\times n}$, we build the array $d[1..n]$ which stores at position $k$ the number of distinct $k\times k$ submatrices of $A$ then we obtain $\delta_{2D}$ as $\max_k d[k]/k^2$. Initially the \textrm{Isuffix Tree} $IST(A)$ of $A$ is built in time $O(n^2)$ \cite{algorithmica/KimNSP11}, then $IST(A)$ is visited in depth first order.
Let $u$ be a node such that the path from the root to $u$ contains 
$|L(u)|$ \textrm{Icharacters}. Let $e$ be an edge outgoing from $u$ labeled with $q_e=\langle i,j,\ell_1,\ell_2 \rangle$ where $\ell_1=|L(u)|+1$. The \textrm{Istring} of a distinct square submatrix is obtained whenever appending an \textrm{Iprefix} of $I_{A_{ij}}[\ell_1,\ell_2]$ to $L(u)$ yields an \textrm{Istring} of odd \textrm{length}. Because the traversing of $e$ {may} yields new square submatrices, $d[\cdot]$ must be updated accordingly. Let $s=\lceil\frac{\ell_1-1}{2}\rceil+1$ and $t=\lceil\frac{\ell_2}{2}\rceil$. Every $d[k]$ with $k\in [s,t]$ should be increased by one: to do this in constant time we set $d[s]=d[s]+1$ and $d[t+1]=d[t+1]-1$ and we assume that each value stored in an entry $d[i]$ is {\em implicitly} propagated to positions $i+1, i+2, \ldots n$: so the $+1$ is propagated from $s$ up to $t$ and the propagation is canceled by the $-1$ added at the position $t+1$.
At the end of the Isuffix tree visit, for each $k\in [1,n-1]$ we set $d[k+1]=d[k+1]+d[k]$ so that $d[k]$ contains the number of distinct $k\times k$ matrices encountered during the visit and we can compute $\delta_{2D}$ as $\max_k d[k]/k^2$.

Note that when leaf {$l_{ij}$} is reached via the edge $e$ with label $q_e=\langle i,j,\ell_1,\ell_2 \rangle$, all the \textrm{Iprefixes} of $I_{A_{ij}}[\ell_1,\ell_2]$ that have an \textrm{Icharacter} which includes some $\$_x$ symbol should not be counted. {The range of well formed \textrm{Iprefixes} can be determined in constant time since it suffices to access one symbol in each of the last two trailing \textrm{Icharacters} of $I_{A_{ij}}[\ell_1,\ell_2]$ to check whether these two contains any $\$_x$.} Since the \textrm{Isuffix Tree} can be constructed and visited in $O(n^2)$ time the overall time and space complexity of the above algorithm is $O(n^2)$. \qed
\end{proof}

Having established that $\deltaDD$ can be computed in linear time, we now study how large can be the gap between the two measures $\gammaDD$ and $\deltaDD$, recalling that by Lemma~\ref{lemma:delta<gamma} it is $\deltaDD \leq \gammaDD$. In~\cite{titt22} Kociumaka et al. establish a separation result between measures $\delta$ and $\gamma$ by showing a family of strings with $\delta=O(1)$ and $\gamma = \Omega(\log n)$. This bound is tight since they also prove that $\gamma = O(\delta\log\frac{n}{\delta})$. 
The next theorem proves that the gap between the two measures in two dimensions is much bigger: $\delta_{2D}$ can be (asymptotically) smaller than $\gamma_{2D}$ up to a $\sqrt{n}$ factor.

\begin{lemma}\label{lemma:O1}
There exists a family of $n\times n$ matrices {on a constant size alphabet} with $\delta_{2D}=O(1)$ and $\gamma_{2D}=\Omega(\sqrt{n})$
\end{lemma}

\begin{proof}\label{lemma:O1proof}
Consider the matrix $M$ of size $n\times n$ where the first row is the string $S$ composed by $\sqrt{n}/2$ consecutive blocks of size $2\sqrt{n}$ each. The $i^{th}$ block $S_i$ with $i=1,\ldots,\sqrt{n}/2$ is the string $1^i 0^{(2\sqrt{n}-i)}$, so $S_i$ contains (from left to right) $i$ initial ones and the remaining positions are zeros. The remaining rows of the matrix are all equals to $\#^n$. Note that for any size $k$ all distinct submatrices start in the first row or are equal to $\#^{(k\times k)}$. Let $\delta_k$ be $d_{k\times k}/k^2$, so that $\deltaDD$ can be rewritten as $\max \{\delta_k \mid k \in [1,n]\}$. We now show that $\delta_k=O(1)$ for all $k$. For $k=1$, we have $\delta_{1} = |\Sigma| = 3$. 
For $k \geq \sqrt{n}$ it is $\delta_k=O(1)$ since $k^2 \geq n$ and there are at most $(n-k+1)+1\leq n$ distinct $k \times k$ submatrices (there are only $(n-k+1)$ possible positions for a submatrix upper left corner on the first row).  
Now consider $\delta_k$ with $k\in [2,\sqrt{n})$. All distinct $k \times k$ submatrices (excluded the $\#^{(k\times k)}$ one) are those having as first row a distinct substring of length $k$ of $S$. We claim all those substrings are included in the language $0^a1^b0^c$ with $a\in[0,k], b\in [0,k-a], c \in [0,k-a-b]$ such that $a+b+c=k$. To see this, note that no substring of length $k<\sqrt{n}$ can contains any two non adjacent (and non empty) groups of ones since there is a group of at least $\sqrt{n}>k$ consecutive zeros between each of them in $S$.
Fixed $k$, to count the strings in $0^a1^b0^c$ is enough to count the possible choices for the starting/ending positions of the middle $1^b$ block: which are $O(k^2)$, then for $k\in [2,\sqrt{n})$, $\delta_k = \frac{O(k^2)}{k^2} = O(1)$. This proves that $\deltaDD = O(1)$.

To estimate $\gammaDD$ consider the $i^{th}$ block on the first row: $S_i= 1^i 0^{(2\sqrt{n}-i)}$. Each $S_i$ with $i=1,\ldots,\sqrt{n}/2$ is a unique occurrence since the sequence $1^i$ occurs only inside blocks $S_j$ with $j\geq i$ which begins with at least $i$ ones, but inside $S_j$ with $j>i$ the sequence $1^i$ is followed by $2\sqrt{n}-j < 2\sqrt{n}-i$ zeros, so the copy of $S_i$ will intersect the $(j+1)^{st}$ block where no leading zeros are present. As a consequence each submatrix $M_i$ of size $2\sqrt{n} \times 2\sqrt{n}$ having $S_i$ as first row is a unique occurrence too. As each $M_i$ do not overlap any other $M_j$ with $j\neq i$ at least $\sqrt{n}/2$ positions are needed in $\Gamma_{2D}$ to cover them. This proves that $\gammaDD=\Omega(\sqrt{n})$. \qed
\end{proof}

Given a set $S$, the worst-case entropy~\cite{titt22} of $S$ defined as $\lceil \log_2 |S| \rceil$ is the minimum number of bits needed to distinguish (and therefore encode) all the elements in $S$. In the following Lemma, we extend the construction of Lemma~\ref{lemma:O1} to define a family $\mathcal{F}$ of matrices with constant $\deltaDD$ and {worst-case entropy} $\Omega(\sqrt{n}\log n)$. 

%This implies that the matrices in $\mathcal{F}$ cannot be all represented in $o(\sqrt{n})$ words of space.}

%$O(\delta \log \frac{n}{\delta})$  $\deltaDD$ is not reachable 

\begin{lemma} \label{lemma:O1bis}
There exists a family of square matrices on a constant size alphabet with $\deltaDD=O(1)$ and worst-case entropy $\Omega(\sqrt{n}\log n)$.
\end{lemma}

\begin{proof}
Consider again the matrix $M$ of Lemma~\ref{lemma:O1}. Each of the $(\sqrt{n}/2)!$ matrices obtained permuting the $\sqrt{n}/2$ blocks $S_i$ on the first row of $M$ has still $\deltaDD=O(1)$. On the other hand, every encoding algorithm to distinguish among these matrices needs at least $\log((\sqrt{n}/2)!)=\Theta(\sqrt{n}\log n)$ bits. \qed
\end{proof}

\subsection{A glimpse on 3D measures}\label{sec:3D}

The measures $\gamma$ and $\delta$ have a natural generalization also for a number of dimensions $D$ greater than~2. 
The study of this generalization is beyond the scope of this paper; however we generalize Lemmas~\ref{lemma:delta<gamma}, \ref{lemma:O1} and~\ref{lemma:O1bis} to the case $D=3$: our results suggest that the gap between $\gamma$ and $\delta$ increases with the number of dimensions. We introduce the following generalization of $\gamma$ and $\delta$ to cubic matrices.

\begin{definition}\label{def:gamma3}
An attractor $\Gamma_{3D}$ for a cubic matrix $C\in \Sigma^{n\times n\times n}$ is a set of positions $\Gamma_{3D} \subseteq \{1,...,n\}^3$ such that any cubic submatrix of $C$ has an occurrence including a position $p \in \Gamma_{3D}$. The measure $\gamma_{3D}(C)$ is defined as the cardinality of a smallest attractor for $C$. 
\end{definition}

\begin{definition} \label{def:delta3}
Given $C\in\Sigma^{n\times n\times n}$, let $d_{3k}(C)$ be the number of distinct $k\times k \times k$ submatrices of $C$, then
\begin{equation*}
    \delta_{3D}(C)= \max \{d_{3k}(C)/{k^3}\colon k \in [1, n]\}
\end{equation*}
\end{definition}

Repeating verbatim the proof of Lemma~\ref{lemma:delta<gamma} we can easily prove that also for $D=3$ it is $\delta_{3D}\leq \gamma_{3D}$:

\begin{lemma}\label{lemma3:delta<gamma3}
$\delta_{3D}(C)\leq \gamma_{3D}(C)$ for any cubic matrix $C\in \Sigma^{n\times n\times n}$.
\end{lemma}

\begin{lemma}\label{lemma:O1_3D}
There exists an infinite family of cubic matrices $C\in\Sigma^{n\times n\times n}$ on a constant size alphabet $\Sigma$ with $\delta_{3D}=O(1)$ and $\gamma_{3D}=\Omega(n^{2/3})$.
\end{lemma}

\begin{proof}
Assume $\sqrt[3]{n}$ is an even integer. For $i=1,\ldots,(\sqrt[3]{n}/2)^2$, let $S_i$ be the binary string of size $2n^{2/3}$ such that $S_i$ contains $i$ initial ones and $2n^{2/3}-i$ final zeros.
Consider the cubic matrix $C\in\Sigma^{n\times n\times n}$ over the alphabet $\Sigma=\{1,0,\#\}$ defined as follows: for any $i$ of the form $i=j(2n^{2/3})+1$ where $j\in \{0,1,\ldots,\sqrt[3]{n}/2-1\}$, $C[i][1..n][1]$ is equal to the string $S_{j(\sqrt[3]{n}/2)+1}\cdot S_{j(\sqrt[3]{n}/2)+2}\,\cdots\, S_{(j+1)(\sqrt[3]{n}/2)}$ and all the other cells of $C$ are filled with the $\#$ symbol. 
Let $\delta_k$ be $d_{3k}/k^3$, so that $\delta_{3D}$ can be rewritten as $\max \{\delta_k \mid k \in [1 , \; n]\}$. Note that all distinct $k \times k\times k$ submatrices (excluded the $\#^{(k\times k\times k)}$ one) could differ only by the symbols laying on the face $C[1..n][1..n][1]$ of $C$, therefore to count them we count how many distinct $k\times k$ submatrices there are in $C[1..n][1..n][1]$.
We show that $\delta_k=O(1)$ for all~$k$. For $k=1$, we have $\delta_{1} = |\Sigma| = 3$. When $k \geq n^{2/3}$ it is $\delta_k=O(1)$ since $k^3 \geq n^2$ and there at most ${(n-k+1)}^2+1\leq n^2$ possible distinct $k\times k$ submatrices in $C[1..n][1..n][1]$.
Now consider $\delta_k$ with $k\in [2,\ldots,n^{2/3})$: since $k<n^{2/3}$, any $k\times k$ submatrix can contain at most one row different from $\#^k$, therefore these submatrices are distinguished by which distinct length-$k$ substrings of two adjacent strings $S_i\cdot S_{i+1}$ they include in one of their $k$ rows. Since at most each of the $O(k^2)$ distinct length-$k$ substrings (see proof of Lemma~\ref{lemma:O1proof}) could appear in one of the $k$ possible rows, $d_{3k}$ is $O(k^3)$ and hence $\delta_k=O(1)$ for any $k$ as claimed. This proves that $\delta_{3D} = O(1)$.

To estimate $\gamma_{3D}$ we note that each $S_i$ is a unique occurrence and therefore each cubic submatrix with edges of size $2n^{2/3}$ having an $S_i$ on the top edge is a unique occurrence too. Since the distinct $S_i$ are $(\sqrt[3]{n}/2)^2$, such cubic matrices are $\Theta(n^{2/3})$ and do not overlap, thus, at least $\Theta(n^{2/3})$ positions are needed in any attractor to cover them and $\gamma_{3D}=\Omega(n^{2/3})$. \qed
\end{proof}

\begin{lemma}
There exists a family of cubic matrices on a constant size alphabet with $\delta_{3D}=O(1)$ and worst-case entropy $\Omega(n^{2/3}\log n)$
\end{lemma}

\begin{proof}
Consider the matrix $C$ of Lemma \ref{lemma:O1_3D}; any permutation of the $\frac{n^{2/3}}{4}$ strings $S_i$ induces a distinct cubic matrix with common measure $\delta_{3D}=O(1)$. By the standard counting argument, there exists at least one matrix in this family which requires $\log((\frac{n^{2/3}}{4})!)=\Omega(n^{2/3}\log n)$ bits to be represented. \qed
\end{proof}

\section{Two-dimensional bidirectional macro schemes}\label{sec:bidi}

In this section we consider the generalization to two dimensions of bidirectional macro schemes. Given an $n\times n$ input matrix $M$, the natural generalization of the one-dimensional macro scheme as described in Section~\ref{sec:notation} is to consider a partition of $M$ into a set of non-overlapping square submatrices such that each submatrix is either an explicit symbol, i.e. a $1\times 1$ submatrix, or is a copy of another submatrix. As in the one-dimensional case, we require that any symbol in $M$ can be determined through a sequence of copy operations. Unfortunately, the above definition is too restrictive, as is proven by the following lemma.

\begin{lemma} \label{lemma:notoverlap}
The number of submatrices in a partition as described above is at least $\log n$ regardless of the content of the input matrix~$M$.    
\end{lemma}

\begin{proof}
The problem of partitioning an $n \times n$ matrix into non-overlapping square submatrices is known as the ``Mrs. Perkins's quilt problem''. Conway~\cite{conway_1964} has shown that, if the 
greatest common divisor of all submatrix sizes is~1, then any partition of an $n\times n$ square contains at least $\log n$ submatrices. The lemma follows observing that, regardless of the content, any macro scheme for the matrix $M$ consists of non-overlapping submatrices and at least one of them must be {an explicit symbol} of size $1\times 1$, so the greatest common divisor of their size is exactly~1. \qed
\end{proof}

The above result shows that considering non-overlapping square partitions imposes a strong ``geometric'' constraint which is independent on the information stored in the input matrix. A possible route for removing such constraint is to consider partitions consisting of rectangular submatrices. However, for uniformity with the measures $\gammaDD$ and $\deltaDD$, we still consider square submatrices, but we relax the assumption that they are non-overlapping. To help the reader in the following we use the term {\em parsing} instead of partition when the submatrices can overlap.

Formally, given an $n\times n$ input matrix $M$, an (overlapping) bidirectional macro scheme consists of a parsing of $M$ in square submatrices, called phrases, such that each phrase $p$ is either a single explicit symbol or there exists another submatrix, called source, from which $p$ is copied. More precisely, we say that the $\ell\times\ell$ submatrix starting at position $(u,v)$ is copied from position $(u+h_1,v+h_2)$ if 
\begin{equation}\label{eq:2Dphrase}
\forall i,j\colon 
\begin{cases}
 u \leq i < u+\ell \\ v \leq j < v+\ell   
\end{cases}
\qquad 
M[i][j] = M[i+h_1][j+h_2]    
\end{equation}
($h_1$ and $h_2$ can be negative). Because of possible phrase overlaps, the same position $(i,j)$ can be contained in different phrases and its value $M[i][j]$ can be copied from different sources. Hence, the whole bidirectional scheme cannot be summarized by a single function~$f$ as in the one-dimensional case (see Sect.~\ref{sec:notation}); instead for each phrase $p$ we define a function $f_p$ whose domain $D_p$ coincides with~$p$. For example, for the phrase $p$ defined by~\eqref{eq:2Dphrase} we have:
$$
D_p = \{(i,j) \mid  u \leq i < u+\ell,\; v \leq j < v+\ell \},\qquad f_p(i,j) = (i+h_1,j+h_2).
$$
In the special case in which a phrase $q$ consists of an explicit symbol, i.e. a $1\times 1$ submatrix $M[i][j]$, the domain of $f_q$ is the singleton $D_q = \{(i,j)\}$, and we define $f_q(i,j) = \bot$.
As for the one-dimensional case, we say that the macro scheme is {\em valid} if it is possible to retrieve the identity of any element of the input matrix by successive copy operations eventually leading to an explicit symbol. Formally, we require that for any position $(i,j)$ {there must be} a sequence of phrases $p_1, \ldots p_k$ such that 
\begin{equation}\label{eq:bot}
f_{p_k}( f_{p_{k-1}}( \cdots f_{p_1}(i,j)\cdots)) = \bot.    
\end{equation}

\begin{example}\label{ex:Gamma}
Let $\Maa$ denote the matrix $n\times n$ containing only the symbol~$a$.  A valid macro scheme for $\Maa$ consisting of an $1\times 1$ phrase and three $(n-1) \times (n-1)$ phrases is the following:
\begin{align*}
    D_0 &= \{(1,1)\}, 
    & D_1 &= \{(i,j)\mid 2 \leq i \leq n,\; 1 \leq j < n \},\;\\
    D_2 &= \{(i,j)\mid 1 \leq i < n,\; 2 \leq j \leq n \},
    & D_3 &= \{(i,j)\mid 2 \leq i \leq n,\ 2 \leq j \leq n \};
\end{align*}
with the corresponding functions:
$$
f_0(1,1) = \bot,\quad f_1(i,j) = (i-1,j),\quad
f_2(i,j) = (i,j-1),\quad f_3(i,j) = (i-1,j-1).
$$
Note that $f_1$ copies the value $(i,j)$ from the position  immediately above, $f_2$ from the position immediately on the left, and $f_3$ from the position adjacent to its upper left corner. Combining these copy operations we can retrieve the content of any entry in~$\Maa$ from the explicit symbol in $(1,1)$. With reference to the rule~\eqref{eq:bot}, we have
$$
f_3(4,2) = (3,1),\quad f_1(3,1) = (2,1),\qquad 
f_1(2,1) = (1,1),\quad f_0(1,1) = \bot.
$$
One can prove that there cannot be a macro scheme with less than four phrases observing that no phrase can include more than one corner of the input matrix.
\end{example}

Note that by Lemma~\ref{lemma:notoverlap} a partition of $\Maa$ into non-overlapping phrases requires at least $\log n$ phrases. Thus, measuring a matrix complexity considering also overlapping phrases appears to be a more natural choice. Note however that the main result of this section (Theorem~\ref{theo:b<delta}) uses only non-overlapping phrases so it is valid in both contexts. 

\begin{definition}\label{def:macros1D}
    For any $n\times n$ matrix $M$, we write $\bDD(M)$, or simply $\bDD$ when $M$ is clear from the context, to denote the minimum number of (possibly overlapping) phrases in a valid two-dimensional bidirectional macro schemes for $M$. 
\end{definition}

As in the one-dimensional case, the measure $\bDD$ is {\em reachable} in the sense that any matrix $M$ can be represented in $O(\bDD(M))$ words. Finding the size of the minimal one-dimensional bidirectional macro scheme is NP-complete, so we conjecture that computing $\bDD$ is NP-complete as well.

According to Definition~\ref{def:macros1D}, for the matrix $\Maa$ of Example~\ref{ex:Gamma} it is $\bDD(\Maa) = 4$. It is easy to see that $\deltaDD(\Maa) = \gammaDD(\Maa) = 1$. Thus, it is $\deltaDD(\Maa) \leq \gammaDD(\Maa) \leq \bDD(\Maa)$ which is the same relationship which holds in the one-dimensional case, where for any input string $S$, $\delta(S) \leq \gamma(S) \leq b(S)$. However, for two dimensions these inequalities are not always valid. Indeed, we now show that $\bDD$ can be significantly smaller than $\deltaDD$ and $\gammaDD$. As a preliminary result we prove that for non-constant alphabets the gap between the measures can be $\Omega(\sqrt{n})$ in favor of~$\bDD$.

\begin{figure*}[t]
    \centering
	\subfloat[]{\includegraphics[scale=0.55]{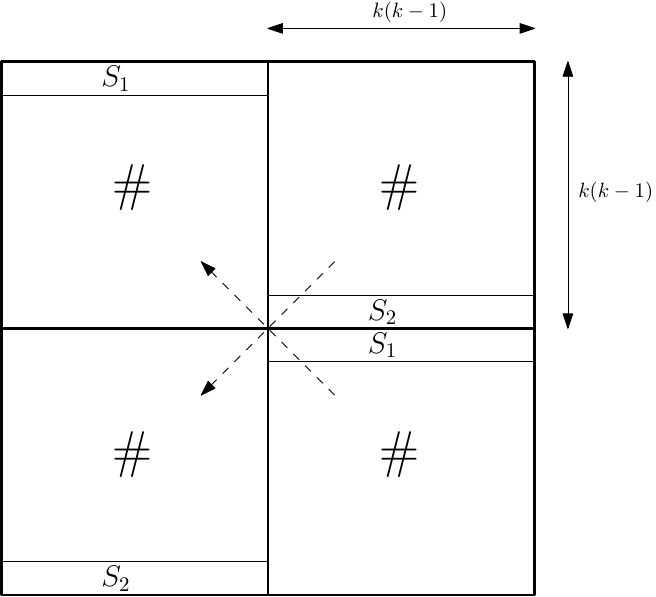}\label{fig:gap_b_y}}\hspace{0.7cm}
	\subfloat[]{\includegraphics[scale=0.55]{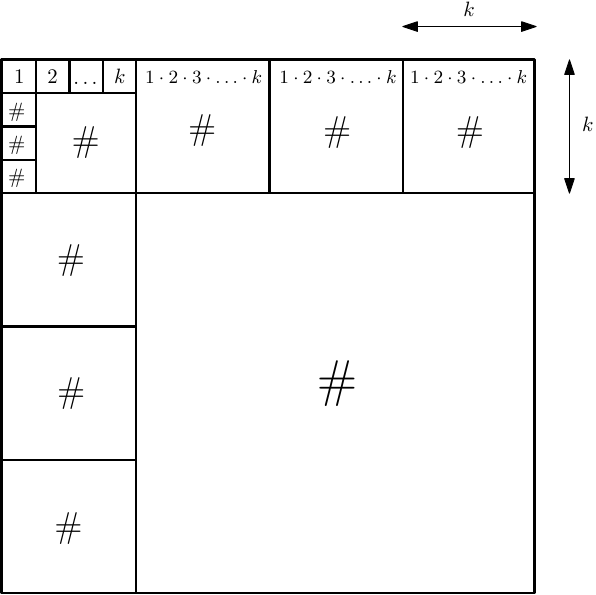}\label{fig:gap_b_y_zoom}} \\
    \caption{The matrix used in the proof of Lemma~\ref{lemma:gap_b_gd_1} (a), and a parsing of its upper left quadrant (b).}
    \label{fig:gap2}
\end{figure*}

\begin{lemma}\label{lemma:gap_b_gd_1}
There exists an infinite family of {$n\times n$} matrices over an alphabet $\Sigma$ of size $\Theta(\sqrt{n})$ such that $\bDD=O(\sqrt{n})$ and $\deltaDD=\Omega(n)$.
\end{lemma}

\begin{proof}
We prove that there exists an infinite family of matrices over an alphabet $\Sigma$ of size $k+1$ with $\bDD=O(k)$ and $\deltaDD=\Omega(k^2)$. Rewriting $k$ in terms of $n$ we get the result as stated above.

Consider the $n\times n$ matrix $M$ with $n=2k(k-1)$ over the alphabet $\Sigma=\{1,\ldots ,k\}\cup \{\#\}$ represented in Figure~\ref{fig:gap_b_y}. If we logically partition $M$ into 4 quadrants of size $k(k-1) \times k(k-1)$, the lower and the upper quadrants on the right are identical to the upper and lower quadrants on the left, respectively, as shown by the dashed arrows. In addition, both $S_1$ and $S_2$ are periodic strings of length $k(k-1)$: $S_1 = (1 2 \cdots k)^{k-1}$ and $S_2 = (2\cdots  k)^{k}$. The remaining cells of $M$ contain the $\#$ symbol.

We now prove that there exist a valid bidirectional macro scheme for $M$ of size $O(k)$. Clearly the right half of $M$ can be parsed in two phrases with sources in the left half. We only show a parsing of the upper left quadrant in $O(k)$ phrases, since the parsing for the lower left quadrant can be obtained in a similar way. 
Consider the partition of the upper left quadrant shown in Figure~\ref{fig:gap_b_y_zoom}.
The $k\times k$ submatrix on the upper left corner is parsed into $2k$ phrases: $2k-1$ explicit symbols on the upper and left border and a single $k-1 \times k-1$ phrase (containing only the symbol \#) which is copied by one position to its left. The $k-2$ phrases of size $k\times k$ on the top are copied from the $k\times k$ submatrix on the upper left corner. The $k-2$ phrases of size $k\times k$ on the right (containing only the symbol \#) are copied by one position above. Finally, the submatrix of size $k(k-2) \times k(k-2)$ on the bottom right corner is copied by one position to its left. Overall, the above macro scheme for $M$ has $O(k)$ phrases, hence $\bDD(M) = O(k)$. Notice also that every entry of $M$ is contained in a single phrase, hence the parsing is non-overlapping. 

To show that $\deltaDD=\Omega(k^2)$, consider the border between the upper right quadrant and the lower right quadrant, where we have a copy of $S_2$ immediately above a copy of $S_1$. We claim that there are not two indices $i, j$, with $1 \leq i< j \leq k(k-1)$ such that $S_2[i]=S_2[j]$ and $S_1[i]=S_1[j]$. To see this, observe that $S_2$ and $S_1$ have periodicity $k-1$ and $k$, respectively, so the hypothesis would imply that $j-i$ be a multiple of both $k-1$ and $k$. Since $k-1$ and $k$ do not have any common factor, $j-i$ should be a nonzero multiple of $k(k-1)$ which is impossible since $j-i<k(k-1)$.  It follows that the $k(k-1)-1$ submatrices of $M$ of size $2\times 2$ having the first row in $S_2$ and the second in $S_1$ are all distinct. Hence $d_{2\times 2}(M) \geq k(k-1)-1$ and $\deltaDD(M) = \Omega(k^2)$ as claimed. \qed
\end{proof}

The above result is not completely satisfactory since the alphabet size is {$k+1=\Theta(\sqrt{n})$}. Next theorem generalizes the above lemma to constant size alphabet: the idea is to replace each symbol $1,2,\ldots,k$ in the matrix $M$ of Lemma~\ref{lemma:gap_b_gd_1} with a different $\ell\times\ell$ binary matrix. Since there are $2^{\ell^2}$ distinct such matrices, increasing $\ell$ will have the same effect as increasing the parameter~$k$ without changing the alphabet.

%Next we extend this result and show that even for a constant size alphabet the gap continues to be notable, albeit reduced by a factor $\Theta(\log^c n)$ with $c=2$ for $\gammaDD$ and $c=5/2$ for $\deltaDD$.

\begin{theorem}\label{theo:b<delta}
There exists an infinite family of matrices over the alphabet $\Sigma=\{0,1,\#\}$ such that $\bDD=O(\sqrt{n}\log n)$, 
$\gammaDD=\Omega(n/\log n)$ and $\deltaDD=\Omega(n/\log^{3/2} n)$.
\end{theorem}

\begin{proof}

For any integer $\ell>0$ consider the set of $k = 2^{\ell^2}$ distinct binary matrices of size $\ell\times\ell$. Let $M'$ denote the matrix over the alphabet $\Sigma=\{0,1,\#\}$ obtained from the matrix $M$ of Lemma~\ref{lemma:gap_b_gd_1} replacing each element with an $\ell\times\ell$ matrix as follows: each symbol $i$ for $i=1,\ldots,k$, is replaced by a distinct $\ell\times\ell$ binary matrix, while the symbol $\#$ is replaced by the   $\ell\times\ell$ matrix $\#^{\ell\times\ell}$. By construction the matrix $M'$ has {side} $n=2\ell k (k-1) = 2\ell 2^{\ell^2} (2^{\ell^2}-1)$. We prove the theorem by estimating $\bDD(M')$ vs $\deltaDD(M')$ and $\gammaDD(M')$.

To upper bound $\bDD(M')$, consider first the submatrix on the top left {corner} of Figure~\ref{fig:gap_b_y_zoom}, which in $M'$ has size $\ell k \times \ell k$. We parse {its first row}, consisting of all the $k=2^{\ell^2}$ binary $\ell\times\ell$ matrices, with $\ell^2 k$ explicit symbols. Next, we parse a single $\#^{\ell\times\ell}$ matrix with $\ell^2$ explicit symbols, and the remaining $k-2$ matrices $\#^{\ell\times\ell}$ as $k-2$ phrases. The remaining submatrix of size $\ell(k-1) \times \ell(k-1)$ containing only the symbol $\#$ is parsed in a single phrase copying its content from one position to its left. Summing up, we parse the top left submatrix of Figure~\ref{fig:gap_b_y_zoom} in $O(\ell^2 k)$ phrases. Reasoning as in the proof of Lemma~\ref{lemma:gap_b_gd_1}, the rest of the $M'$ matrix can be parsed in $O(k)$ phrases, hence we have $\bDD = O(\ell^2 k)$. Since {$n =\Theta(\ell k^2) = \Theta(\ell 4^{\ell^2})$}, we have $k=O(\sqrt{n})$ and $\ell = O(\sqrt{\log n})$ {therefore} $\bDD(M') = O(\sqrt{n}\log n)$.

To obtain a lower bound on $\deltaDD(M')$ we reason as in Lemma~\ref{lemma:gap_b_gd_1} and we count the number of $2\ell \times 2\ell$ binary matrices of $M'$ which are inside the submatrix $R$ consisting of a copy of $S_2$ immediately above a copy of $S_1$ (in $M'$ such submatrix has size $2\ell \times \ell k (k-1)$). Reasoning as in the proof of Lemma~\ref{lemma:gap_b_gd_1} we see that there are at least $k(k-1) -1  = \Omega(k^2)$ distinct $2\ell\times 2\ell$ binary submatrices in $R$, hence $\deltaDD(M') = \Omega(k^2/\ell^2) = \Omega(n/\ell^3) = \Omega(n/(\log n)^{3/2})$. 

Finally, to lower bound $\gammaDD(M')$ we consider again the $\Omega(k^2)$ distinct binary $2\ell\times 2\ell$ matrices inside $R$. 
{We observe that these submatrices only occur inside $R$, as there are no other $2\ell\times 2\ell$ binary submatrices outside of $R$. Therefore, each attractor position covering any of these distinct submatrices must belong to $R$ and is contained in at most $2\ell$ of such submatrices.} We conclude that any attractor must have size {$\Omega(k^2/\ell) = \Omega(n/\ell^2) = \Omega(n/\log n)$ }, as claimed. \qed
\end{proof}

Note that also in the proof of Theorem~\ref{theo:b<delta} the phrases in the proposed bidirectional macro scheme are non-overlapping. In Section~\ref{sec:2DBT}, we will use the two-dimensional block tree to build (not necessarily minimal) macro schemes, and we will provide general bounds for $\bDD(M)$ in terms of  $\deltaDD(M)$ and $\gammaDD(M)$.

\section{Space Bounds for Two-Dimensional Block Trees}\label{sec:2DBT}

Brisaboa et al.~\cite{2dbt} generalized the Block Tree concept~\cite{jcss2020} to two dimensional data providing a compressed representation for discrete repetitive matrices that offers direct access to any compressed submatrix in logarithmic time. Given a matrix $M\in\Sigma^{n\times n}$ and an integer parameter $k>1$, assume first for simplicity that $n$ if a power of $k$, i.e. $n=k^\alpha$. We build the Two-dimensional Block Tree (\BTDD from now on) as follows. The root of the tree at level $\ell=0$ represents the whole matrix $M$. In the next level we store $k^2$ nodes, each node represents a distinct non overlapping $n/k \times n/k$ submatrix of $M$. In general, to build the level $\ell$ of the \BTDD we recursively partition (some of) the submatrices represented at level $\ell-1$ into $k^2$ submatrices of size $n/k^\ell \times n/k^\ell$ and for each of these smaller submatrices we store a corresponding descending node in the level $\ell$ of the \BTDD. In the following we call a {\em level-$\ell$ block} (or block at level $\ell$), any $n/k^\ell \times n/k^\ell$ submatrix of $M$ whose upper left corner is an entry of the form $M[1+i (n/k^\ell)][1+j (n/k^\ell)]$ with $i$ and $j$ non-negative integers. A node at level $\ell$ of the \BTDD represents a level-$\ell$ block; however not all blocks are necessarily represented in the tree: the \BTDD prunes redundant subtrees, meaning that if a block occurs elsewhere the corresponding node is pruned and replaced by pointers. Note that, since not necessarily all nodes in one level are expanded in the next, the resulting structure is a tree which can have leaves at different heights. Our pruning strategy consists in considering as candidate for pruning every node whose corresponding block $B$ appears in a submatrix $O$ which precedes $B$ in row major order. Formally, we make use of the concept of {\em first occurrence}:

\begin{definition}\label{def:first:occ}
    We say that a non-empty submatrix $O = M[a..a+x][b..b+y]$ is a {\em first occurrence} if there is not another submatrix $M[c..c+x][d..d+y]$ with the same content of $O$ which precedes~$O$ in row major order, that is, with $c<a$ or $(c=a) \wedge (d<b)$. Note that if a submatrix $V$ contains $O$ which is a first occurrence, then also $V$ is a first occurrence.
\end{definition}

Whenever a level-$\ell$ block is a candidate for pruning we must also ensure that its corresponding first occurrence $O$ is available in the \BTDD. To this end, we consider all the submatrices of size $2n/k^\ell \times 2n/k^\ell$ made of 4 adjacent level-$\ell$ blocks, and we refer to them as {\em superblocks}. We call {\em block-marker} any superblock $D$ which is a first occurrence according to the above definition.

\begin{figure}[t]
    \centering
    \includegraphics[scale=0.40]{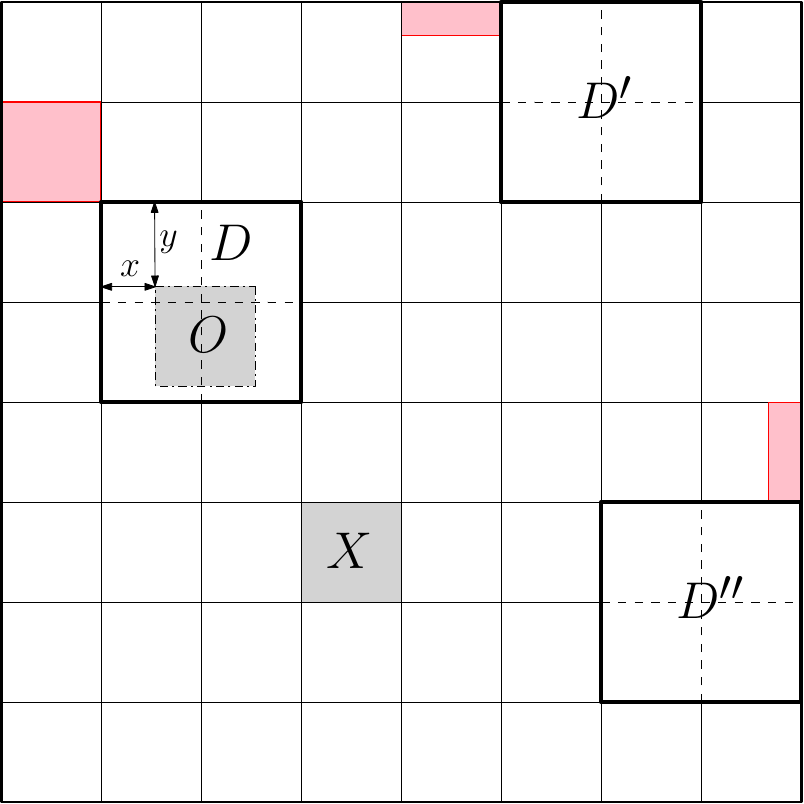}
    \caption{A block $X$ and its first occurrence $O$ in row-major order. If $X$ is not marked, its node $X_v$ in the \BTDD will point to the four blocks overlapping $O$ and will store the offset $\langle x,y \rangle$ of $O$ with respect to the block including $O$'s top left corner. The pointed blocks are marked since $D$ includes $O$ and therefore is a block-marker.
    With reference to the proof of Lemma~\ref{lemma:2DBTnodes}, $D$ is a type 3 block-marker: by considering every entry in the red block as an upper left corner we obtain $k^{2\ell}$ distinct $3k^\ell\times 3k^\ell$ submatrices containing $D$. The type 2 block-marker $D'$ borders the upper edge; by considering the $k^\ell$ entries in the first row marked in red we obtain $k^{\ell}$ distinct $3k^\ell\times 3k^\ell$ submatrices containing $D'$. We also show a type 2 block-marker $D''$ bordering the right edge; we obtain $k^{\ell}$ distinct $3k^\ell\times 3k^\ell$ submatrices containing $D''$ by considering the $3k^\ell\times 3k^\ell$ submatrices with the upper {\em right} corner in the portion of the last column marked in red.}
    \label{fig:superblock}
\end{figure}

At any level $\ell$ of the \BTDD, the four level-$\ell$ blocks included in a block-marker are {\em marked}; conversely a block $B$ is {\em unmarked} if none of the (up to 4) level-$\ell$ superblocks including $B$ in a corner (they are less than 4 when $B$ touches one of the matrix borders) is a block-marker. 
In the following we call $B_v$ the block represented by node $v$ in the \BTDD, and we say that a node $v$ is marked (unmarked) if and only if $B_v$ is marked (unmarked), therefore, the previous marking rule for blocks can be applied equivalently to the nodes of the \BTDD. 
All level $\ell$ marked nodes are expanded at level $\ell+1$, instead all the other (unmarked) nodes in the level are the level-$\ell$ leaves of the \BTDD and store additional information that we use to retrieve the content of the corresponding pruned subtree: an unmarked node $u$ points to the (at most four) marked nodes in the same level whose corresponding blocks overlap the first occurrence $O$ of the (unmarked) block $B_u$. In addition to the horizontal pointers, $u$ stores the relative offset $\langle x,y\rangle$ of $O$ inside the marked block where $O$ starts (see Figure~\ref{fig:superblock}). 
The splitting process ends when explicitly storing blocks is more convenient than storing pointers to marked blocks, i.e. when the block size is $\Theta(\log_\sigma n)$. The resulting tree-shaped data structure has height $h=O(\log_k (n\sqrt{\frac{\log \sigma}{\log n}}))$. The correctness of the marking scheme is proven by the following lemma.    

% We nonetheless consider this variant since it is the one used in every implementation (see also~\cite{jcss2020}).
%Repeating arguments 1) and 3) shows that any unmarked node always points to nodes which are marked and exist in the same level since none of their ancestors has been pruned in a previous level. 

\begin{lemma}\label{lem:our:marking}
    An unmarked node points to nodes which are marked and exist at the same level since none of their ancestors has been pruned in a previous level.
\end{lemma}

\begin{proof}
We show that the proposed marking scheme guarantees that: 1) if a level-$\ell$ block $B$ overlaps the first occurrence $O$ of any $a\times b$ submatrix with both $a,b \leq (n/k^\ell)$ then $B$ is marked,
2) given a level-$\ell$ unmarked block $X$ the (up to four) blocks intersecting the first occurrence $O_X$ of $X$ are marked, 3) any level-$\ell$ marked block $B$ descends from a marked block at level~$\ell-1$.

Property 1) holds because at least one of the superblocks which include both $B$ and~$O$ is a block-marker, otherwise $O$ would not be a first occurrence. Property 2) follows from 1) applying it to $O_X$. To see why 3) holds, note that if $B$ is marked then there exists a block-marker $D$ which includes $B$. Since $D$ includes $B$, $D$ overlaps the parent block $\mathit{parent}(B)$ of $B$ which has size $n/k^{\ell-1} \times n/k^{\ell-1}$, therefore by property 1) and by the fact that $2n/k^\ell \leq n/k^{\ell-1}$ we conclude that $\mathit{parent}(B)$ is marked. Applying the above argument to all levels proves that none of $B$'s ancestors is pruned as claimed. \qed
\end{proof}

If $n$ is not a power of $k$ the blocks on the right and bottom borders can have rectangular shape. Formally, let $n'=k^{\lceil\log_k n\rceil}$, $a_\ell = n'/k^\ell$ and $b_\ell = n \bmod a_\ell$. 
Then, at level $\ell$ blocks still start in positions of the form $(1+i a_\ell,1+j a_\ell)$ with $i,j$ integers, but the block including $M[n][n]$ is a square of side $b_\ell$ while the remaining blocks which intersect the last column/row of $M$ are rectangles of size respectively $(a_\ell \times b_\ell)$ and $(b_\ell \times a_\ell)$. As a consequence, the superblock on the bottom right corner is a square of side $a_\ell + b_\ell$, while the other superblocks that border the bottom and right edges of $M$ are rectangles respectively of size $(a_\ell+b_\ell) \times 2a_\ell$ and $2a_\ell \times (a_\ell+b_\ell)$. Apart from the shape of such superblocks the construction does not change, i.e. if a superblock is a first occurrence then the four blocks it contains are marked and the corresponding nodes are split and expanded at the next level. The crucial point is that Lemma~\ref{lem:our:marking} still holds since it does not rely on the shape of the blocks.

Our marking strategy, which directly extends the one proposed for the one dimensional block tree \cite{jcss2020}, is slightly different than the one in~\cite{2dbt} in which if a submatrix is pruned at some level its content is seen as all $0$s in the subsequent levels. This approach removes the issue of possibly pointing to pruned nodes, but makes it difficult to estimate the number of marked nodes in terms of the matrix content, which is our next objective. As a preliminary result we bound the number of marked nodes at each level. 

\begin{lemma}\label{lemma:2DBTnodes}
The number of marked nodes in any level of a \textit{2D block tree} is $O(\deltaDD + \sqrt{n\deltaDD})$.
\end{lemma}
\begin{proof}
Assume first for simplicity that $n$ if a power of $k$. To ease the proof we number the levels backwards, that is we say that a node is at level $\ell$ if its height in the \BTDD is $\ell$, similarly a block of level $\ell$ is a $k^\ell \times k^\ell$ submatrix of $M$ whose upper left corner is an entry of the form $M[1+\lambda k^\ell,1+\mu k^\ell]$ with $\lambda, \mu$ positive integers. We note that the number of marked nodes at any given level $\ell$ of the \BTDD is equal to the number of marked blocks at the same level. By construction, a level-$\ell$ block is marked if and only if it is contained in a superblock which is a block-marker for the level $\ell$. Hence, if $u$ is the number of $2k^\ell \times 2k^\ell$ block-markers, the number of marked blocks at level $\ell$ is at most~$4u$. To bound the number of block-markers $u$, we partition them into $3$ types: 1) those on a corner of $M$ i.e. which include one of the entries $M[1][1], M[1][n], M[n][1]$ or $M[n][n]$, these are at most four; 2) those not on a corner but including an entry in the first/last row/column; 3) those not including any entry in the first/last row/column. Let $u_i$ be the number of block-markers of type~$i$: clearly $u = u_1+u_2+u_3=O(u_2+u_3)$. 

Given a block-marker $D$ of the third type, we observe that $D$ is included into $k^{2\ell}$ distinct $3k^\ell \times 3k^\ell$ submatrices starting {at any position in the $k^\ell \times k^\ell$ block} touching the top left corner of $D$ {(see Figure~\ref{fig:superblock})}. Summing over all type 3 block-markers, we have a total of $u_3 k^{2\ell}$ submatrices of size $3k^\ell \times 3k^\ell$ starting in distinct positions inside $M$. These submatrices are distinct: each submatrix contains a block-marker $D$ by construction; if two of those matrices were equal we would have two first occurrences of $D$ starting in different positions which is impossible. 
Since by definition the number of distinct submatrices of size $3k^\ell\times3k^\ell$ is at most $(3k^{\ell})^2\deltaDD$ we have 
$$
u_3 k^{2\ell} \leq 9k^{2\ell}\deltaDD \quad\Longrightarrow\quad u_3 \leq 9\deltaDD
$$
Consider now a block-marker $D'$ of the second type bordering the upper edge of $M$ (the other 3 cases are treated similarly, see the block-marker $D''$ in Figure~\ref{fig:superblock}). Any $3k^\ell \times 3k^\ell$ matrix which starts in the same row as $D'$, but in any of the $k^\ell$ columns preceding $D'$ is distinct by the same argument presented before (see again Figure~\ref{fig:superblock}). Reasoning as above we find $u_2 k^\ell$ distinct $3k^\ell \times 3k^\ell$ matrices which implies $u_2 \leq 9k^\ell\deltaDD$. Since it is also $u_2< 4(n/k^\ell)$, we have $u_2\leq \min(9k^\ell\deltaDD,4 n/k^\ell)=O(\sqrt{n\deltaDD})$. We conclude that the number of marked blocks at any level is $O(u) = O(u_1+u_2+u_3)=O(\deltaDD + \sqrt{n\deltaDD})$. 

\begin{figure}[t]
    \centering
    \includegraphics[scale=0.40]{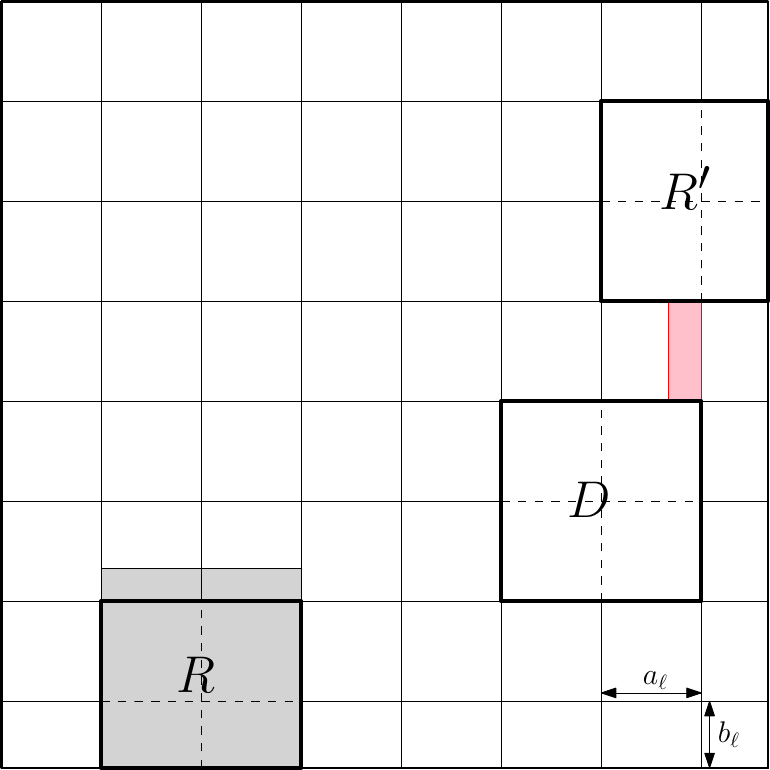}
    \caption{Possible partitioning of a matrix whose size is not a power of $k$: we see that along the right and bottom margin there are rectangular blocks of size $a_\ell \times b_\ell$ or $b_\ell \times a_\ell$. $R$ and $R'$ are rectangular superblocks, if they are a first occurrence the four blocks they contain are marked. With reference to the proof of Lemma~\ref{lemma:2DBTnodes}, the block-marker $D$ is adjacent to rectangular blocks, so we cannot guarantee that there are $k^{2\ell}$ distinct $3k^\ell\times 3k^\ell$ submatrices containing $D$. However, we obtain $k^{\ell}$ distinct $3k^\ell\times 3k^\ell$ submatrices containing $D$ by considering the $3k^\ell\times 3k^\ell$ submatrices with the upper {right} corner in the area marked in red.
    }\label{fig:superblockR}
\end{figure}

If $n$ is not a power of $k$ the result continues to hold with only some minor changes. The superblocks intersecting the last column/row of $M$ now have rectangular shape since they contains rectangular blocks (see Fig.~\ref{fig:superblockR}). However, the number of such superblocks which are block-markers can still be bounded with the same argument used for type 2 block-markers. The only other difference is that for the type 3 block-markers which are {\em adjacent} to a rectangular block, like $D$ in Fig.~\ref{fig:superblockR}, we can no longer guarantee that they are included into $k^{2\ell}$ distinct $3k^\ell \times 3k^\ell$ submatrices. However, since such block-markers are $O(n/k^\ell)$ in total and they are included into $k^{\ell}$ distinct $3k^\ell \times 3k^\ell$ submatrices, their number can be bounded as for type 2 block-markers, so the overall bound on the total number of marked nodes at each \BTDD level does not change. \qed
\end{proof}

%prima era: \log \frac{n \log \sigma}{\sqrt{\deltaDD} \log n}
%$h=O(\log_k (n\sqrt{\frac{\log \sigma}{{\deltaDD}\log n}}))$

\begin{theorem}\label{theorem:2DBTspace}
The \BTDD built on a matrix $M\in\Sigma^{n\times n}$ takes up $O((\deltaDD + \sqrt{n\deltaDD})\log (n\sqrt{\frac{\log \sigma}{\deltaDD \log n}}))$ space. This space is worst-case optimal within an $O(\log n)$ multiplicative factor.
\end{theorem}

\begin{proof}
The \BTDD as described at the beginning of the section has height $O(\log_k (n\sqrt{\frac{\log \sigma}{\log n}}))$. We reduce such height by using, at the level immediately below the root, blocks of size $k^\beta \times k^\beta$ where $\beta=\lfloor \log_k (n/\sqrt{\deltaDD})\rfloor$ is the largest integer such that $k^\beta \leq n/\sqrt{\deltaDD}$. As a result, the height of the tree is reduced to $O(\log_k (n\sqrt{\frac{\log \sigma}{\deltaDD \log n}}))$. Note that even with the above initial splitting, at any level of the \BTDD rectangular blocks can only appear on the right and bottom edges of the input matrix; in addition the number of blocks at the first level is $\lceil n/k^\beta \rceil^2 = \lceil kn/k^{\beta+1} \rceil^2$ which is  $O(\deltaDD)$ since $n/\sqrt{\deltaDD} < k^{\beta+1}$. These two facts show that, even considering the above initial partitioning, the bound of Lemma~\ref{lemma:2DBTnodes} still applies, therefore, summing over all the levels of the lowered tree we get an overall number of marked nodes $m=O((\deltaDD + \sqrt{n\deltaDD})\log_k (n\sqrt{\frac{\log \sigma}{\deltaDD \log n}}))$. The first part of the theorem follows because each marked node produces $O(1)$ unmarked nodes on the next level since $k=O(1)$.

To prove the worst-case quasi-optimality let $\mathcal{F}_n$ denote the set of matrices introduced in the proof of Lemma~\ref{lemma:O1bis}.  For such matrices $\deltaDD=O(1)$, and from Lemma~\ref{lemma:O1bis}, for any coder $C:\mathcal{F}_n\rightarrow{\{0,1\}}^*$ representing all the matrices in $\mathcal{F}_n$, there exist a matrix $W$ such that $|C(W)|=\Omega(\sqrt{n}\log n)$ bits. The result follows since for any matrix in $\mathcal{F}_n$ the \BTDD takes $O(\sqrt{n}\log n)$ words and therefore $O(\sqrt{n}\log^2 n)$ bits of space. \qed
\end{proof}

In~\cite{titt22} it is shown that the one-dimensional Block Tree is worst case optimal in terms of the measure $\delta$, so in this respect Theorem~\ref{theorem:2DBTspace} is not completely satisfactory. Unfortunately,  the following result shows that the bound in Lemma~\ref{lemma:2DBTnodes} cannot be substantially improved at least when $\deltaDD = O(1)$. Since the proof of Lemma~\ref{lemma:2DBTnodes} shows that the number of marked blocks {\em at the interior} of the matrix, i.e. deriving from type 3 block-markers, is bounded by $O(\deltaDD)$, to establish our result we consider a family of matrices that have a hard to compress first row.

\begin{lemma}
There exists an infinite family of matrices $M \in \Sigma^{n\times n}$ over a constant size alphabet $\Sigma$ with $\deltaDD(M)=O(1)$, such that the \BTDD for $M$ has $\Omega(\sqrt{n})$ marked nodes on a single level.
\end{lemma}

\begin{proof}
Let $M$ be the matrix of Lemma~\ref{lemma:O1} with $n = k^{2\alpha}$ so that $n$ is both a power of $k$ and a perfect square. We have already proven that $\delta_{2D}(M)=O(1)$. Consider the \textit{2D-BT} built on $M$: note that for block size larger than $4\sqrt{n}\times 4\sqrt{n}$ each block on the upper edge of $M$ includes entirely in its first row at least one of the strings $S_i$ of the form $1^i 0^{(2\sqrt{n}-i)}$ composing $S$. Since each $S_i$ is unique, any of those blocks is the first occurrence of its content and hence marked. In particular, this applies at the blocks at level $\ell^* = \alpha - \lceil \log_k 4 \rceil$ which have size $k^{2\alpha-\ell^*} \times k^{2\alpha-\ell^*}$ where $k^{2\alpha-\ell^*} = k^{\lceil \log_k 4 \rceil} k^\alpha \geq 4\sqrt{n}$. The blocks at level $\ell^*$ on the upper edge of $M$ are $k^{\ell^*}=k^{\alpha - \lceil \log_k 4 \rceil} = \Theta(k^\alpha) = \Theta(\sqrt{n})$; since they are all marked the corresponding level of the \BTDD has $\Theta(\sqrt{n})$ nodes. \qed
\end{proof}

In~\cite{NAVARRO201941} the authors introduced a variant of the one-dimensional block tree, called $\Gamma$-tree, in which, given a not necessarily minimum string attractor $\Gamma$, the marked nodes at each level are those close to an attractor position. The $\Gamma$-tree is then enriched with additional information that makes it a compressed full text index using $O(\gamma \log(n/\gamma))$ space where $\gamma = |\Gamma|$ is the size of the string attractor. Following the ideas from~\cite{NAVARRO201941}, we now show how to modify the construction of the \BTDD assuming we have available a, not necessarily minimum, {\em 2D-attractor}~${\Gamma_{2D}}$~(see Definition~\ref{def:gamma}).

To simplify the explanation, we initially assume that $n=k^\alpha$ for some $\alpha>0$. Given a matrix $M\in\Sigma^{n\times n}$ and an attractor ${\Gamma_{2D}}$ for $M$, the splitting process is unchanged but we change the marking scheme: at level $\ell$ we mark any node $u$ whose {$n/k^\ell \times n/k^\ell$} block $B_u$ includes a position $p\in {\Gamma_{2D}}$ and the nodes of the (up to eight) blocks adjacent to $B_u$; the remaining nodes are unmarked. An unmarked node $u$ points to an occurrence $O$, of its block $B_u$, which includes an attractor position: $u$ stores at most four horizontal pointers to the blocks on the same level which overlaps $O$ and the relative offset of $O$ within the level-$\ell$ block which includes its top-left corner. The claimed occurrence $O$ which crosses a position $p \in {\Gamma_{2D}}$ necessarily exists otherwise ${\Gamma_{2D}}$ would not be a valid attractor for $M$, furthermore all the level-$\ell$ blocks intersecting $O$ are ensured to be marked as they contain $p$ or are adjacent to a block containing $p$. It is easy to see that the ancestors of a marked node are also marked so the above marking scheme is correct.

In the general case in which $n$ is not a power of $k$, at each level there can be rectangular blocks along the right and bottom border of $M$ (see Figure~\ref{fig:superblockR}).
We still use the same marking rule, but we notice that a rectangular unmarked block is not guaranteed to have an occurrence which contains a position $p\in{\Gamma_{2D}}$. We solve this problem considering, for each rectangular unmarked block $R$, a square submatrix $S$ of size $n/k^\ell \times n/k^\ell$ which includes $R$. The square submatrix $S$ has by definition another occurrence $S'$ which includes a position $p \in {\Gamma_{2D}}$. Therefore, there exists an occurrence $R'$ of $R$ which is included in $S'$. We note that {$R'$ overlaps} only blocks that are marked hence we can safely set up to 4 pointers from the node $u$ representing $R$ to the level-$\ell$ blocks which overlaps $R'$, and store the offset of $R'$ with respect to the block where $R'$ starts. 

\begin{theorem}\label{sec:GTree}
Given a matrix $M\in\Sigma^{n\times n}$ and a two-dimensional attractor ${\Gamma_{2D}}$ for $M$ of size~$\lambda$, the \BTDD built using ${\Gamma_{2D}}$ takes $O(\lambda \log (n\sqrt{\frac{\log \sigma}{\lambda \log n}}))$ space.
\end{theorem}
\begin{proof}
Each position $p\in {\Gamma_{2D}}$ marks at most $9$ distinct blocks per level: the block $B$ including $p$ and the (up to) eight blocks adjacent to $B$, hence, the number of marked blocks per level is at most $9\lambda$. 
As in the proof of Theorem~\ref{theorem:2DBTspace}, we initially divide $M$ into blocks of size $k^\beta \times k^\beta$ with $\beta=\lfloor \log_k (n/\sqrt{\lambda})\rfloor$ so that the root has $O(\lambda)$ children. 
As a result, we get a shallower tree of height $O(\log_k (n\sqrt{\frac{\log \sigma}{\lambda \log n}}))$. Therefore the overall number of marked nodes in the \BTDD is~$O(\lambda \log_k (n\sqrt{\frac{\log \sigma}{\lambda \log n}}))$ and since any marked node produces at most $k^2$ nodes on the next level, being $k=O(1)$ the \BTDD takes $O(\lambda \log (n\sqrt{\frac{\log \sigma}{\lambda \log n}}))$ space. \qed
\end{proof}

% \noindent
Using the \BTDD the following two lemmas bound the measure $\bDD(M)$ in terms of $\gammaDD(M)$ and $\deltaDD(M)$.

\begin{lemma}\label{sec:b_ub}
For any matrix $M\in\Sigma^{n\times n}$ it is $\bDD = O(\gammaDD\log \frac{n}{\sqrt{\gammaDD}})$.
\end{lemma}

\begin{proof}\label{sec:b_ub_proof}
We prove the lemma by showing that the attractor-based \BTDD of Theorem~\ref{sec:GTree}, built using an optimal two-dimensional attractor of size $\gammaDD$, induces a valid bidirectional macro scheme with the claimed number of phrases.
To see why, assume for simplicity that $n$ is a power of $k$ and consider the \BTDD of  height $\log_k n$ in which the root has $k^2$ children and the deepest leaves have size $1\times 1$.
%described in Theorem~\ref{sec:GTree} without optimizations, that is, the tree built down to the leaves of size $1\times 1$ and with second-level blocks of size $n/k \times n/k$ regardless of $\gammaDD$. 
We obtain a valid non-overlapping bidirectional macro scheme by taking the $1\times 1$ submatrices corresponding to the deepest leaves as explicit symbols. The other phrases are the submatrices represented by the remaining leaves, which are copied from the positions reached by their associated horizontal pointers. Since the submatrices corresponding to the leaves of the tree are a partition of $M$, each position in $M$ belongs to exactly one phrase. This observation, together with the fact that we can access any symbol of the input matrix using the \BTDD, shows that the resulting macro scheme is non-overlapping and valid. The number of phrases equals the number of leaves in the \BTDD constructed without any height optimization, which is $O(\gammaDD \log n)$ (see Theorem~\ref{sec:GTree}).

Suppose now $n$ is not necessarily a power of $k$, and consider the \BTDD of Theorem~\ref{sec:GTree} with $M$ partitioned into $O(\gammaDD)$ submatrices at the top level, but with the bottom level leaves of size~$1\times 1$ (the height therefore is $\log \frac{n}{\sqrt{\gammaDD}}$). Since in a bidirectional scheme phrases must be square and we now have rectangular blocks, we need to adjust the above construction. 
Reasoning as in the description immediately before  Theorem~\ref{sec:GTree}, for every rectangular unmarked block $R$ we introduce a phrase $p_R$ by selecting the minimum size square submatrix $S$ that contains~$R$. The source of $p_R$ is the occurrence $S'$ of $S$ that we considered during the initialisation of the horizontal pointers of $R$. It is easy to see that the resulting macro scheme is valid and that it can contain overlapping phrases. By Theorem~\ref{sec:GTree} the number of phrases is $O(\gammaDD\log \frac{n}{\sqrt{\gammaDD}})$ as claimed. \qed
\end{proof}

\begin{lemma} \label{lem:b_delta}
For every matrix $M\in\Sigma^{n\times n}$ it is $\bDD = O((\deltaDD+\sqrt{n \deltaDD}) \log n)$.
\end{lemma}

\begin{proof}
We preliminary observe that we can restrict our attention to the case in which the matrix size is a power of $k$. Indeed, if $n$ is not a power of $k$ we can cover $M$ with $k^2$ square submatrices $M_i$ of size $k^\beta \times k^\beta$ with $\beta=\lfloor \log_k n\rfloor$ (admitting overlapping among them) and we get a valid macro scheme for $M$ by taking the union of $k^2$ macro schemes for the matrices $M_i$.  For each $M_i$ we consider the \BTDD of Theorem~\ref{theorem:2DBTspace} but without optimizations, that is, we consider a tree whose blocks have size $n/k^\ell \times n/k^\ell$ independently of $\deltaDD$, and whose leaves at the deepest level have size $1\times 1$. We build a two-dimensional block tree $T_i$ for each submatrix $M_i$ and we construct a bidirectional macro scheme from each $T_i$ as we do in Lemma~\ref{sec:b_ub_proof}: we take as explicit symbols all the {$1\times 1$ } leaves of $T_i$, and the submatrices corresponding to {the remaining} unmarked leaves are copied phrases. Each bidirectional scheme obtained in this way is a valid non-overlapping scheme for the corresponding $M_i$ because the number of rows/columns in each matrix is a power of $k$. According to Theorem~\ref{theorem:2DBTspace}, each induced bidirectional macro scheme has size $O((\deltaDD(M_i) + \sqrt{n\deltaDD(M_i)})\log n)$, which can be upper bounded by $O((\deltaDD(M) + \sqrt{n\deltaDD(M)})\log n)$ by the monotonicity of $\deltaDD$. The union of all these non overlapping schemes for the $M_i$'s constitutes a valid overlapping scheme for $M$ of $O(k^2(\deltaDD(M) + \sqrt{n\deltaDD(M)})\log n)$ phrases. The lemma follows since $k=O(1)$. \qed
\end{proof}

\section{Two-Dimensional Block Trees construction}\label{sec:2DBT-con}

In~\cite[Sect.~4]{2dbt} the authors present a construction algorithm for their \BTDD based on Karp-Rabin fingerprints extended to two dimensions following the ideas of Bird~\cite{BIRD1977168} and Baker~\cite{BAKER1978}. The cost of this solution is $\Theta(n^3+pn^4)$ where $p$ is the probability of two fingerprints matching (either because of equality of submatrices or because of a collision). For binary matrices, the authors also show how to take advantage of the sparsity of the matrix and compute the \BTDD in $O(mn\log n +pm^2\log_{k^2} (n^2/m))$ time, where $m$ is the number of nonzero elements. 

In this section we present a deterministic $O(n^2)$ algorithm for the construction of the \BTDD described in Section~\ref{sec:2DBT} which works for arbitrary matrices. The main difficulties of the construction are: 1) determining which blocks are marked and 2) for each non marked block determining the position of its first occurrence in $M$. We solve both problems using the Isuffix tree data structure $IST(M)$ reviewed in Section~\ref{sec:delta2D}. 
To simplify the notation, in the following we assume that $n$ is a power of $k$: the general case is analogous and discussed in Appendix~\ref{app:nopow}.  Also to simplify the notation, we stop the construction only when the blocks have size $1\times 1$, so the \BTDD has height $\lmax=\log_k n$. In addition, we number tree levels backwards, that is, we call level $\ell$ the one whose blocks have size $k^\ell \times k^\ell$. We say that a position $(i,j)$ in $M$ is $\ell$-aligned if it is a starting position for a level-$\ell$ block, that is, if both $i-1$ and $j-1$ are integer multiples of $k^\ell$. Note that if a position is $\ell$-aligned then it is $\ell'$-aligned for any $\ell'$  with $0 \leq \ell'\leq \ell$.

Given an $n \times n$ matrix $M$, the Isuffix tree $IST(M)$ has $n^2$ leaves: leaf $l_{ij}$ represents the largest square submatrix of $M$ with upper left corner at position $(i,j)$. Leaves are ordered left to right according to the lexicographic order of the associated IString (see Fig.~\ref{fig:M2}) and we assume we have an array $X[1..n^2]$ such that $X[p]$ stores a pointer to the $p$-th leaf.
To construct the \BTDD we enrich $IST(M)$ with some additional information. First, we build an array $R[1..n^2]$, and its inverse $R^{-1}[1..n^2]$, such that $R[p]$ stores the rank in row major order of the $p^{th}$ leaf. Hence, if $l_{ij}$ is the $p$-th leaf then  $R[p]=n(i-1)+j$ and $R^{-1}[n(i-1)+j] = p$. Clearly $R$ and $R^{-1}$ can be initialized in linear time with a visit of $IST(M)$. During the same visit we also store in each internal node $v$ the range $[s_v,e_v]$ of leaves indices descending from $v$. After constructing $R$ we equip it with a data structure supporting constant time $RMQ$ queries on it. Such data structure can be built in linear time and uses $4n^2+O(\sqrt{n^2\log n}) = \Theta(n^2)$ bits of space~\cite{rmqFischerJohannes}. The array $R^{-1}$ instead is used to find the leaf $l_{ij} = X[R^{-1}[n(i-1)+j]]$ corresponding to an arbitrary matrix position $(i,j)$, an operation that in the following we tacitly assume we can do in $O(1)$ time. 

Finally, we augment $IST(M)$ with some additional pointers. If position $(i,j)$ is {$\ell$-aligned}, that is, there exists a level-$\ell$ block $B$ starting at $(i,j)$, we store a {\em backward pointer} from the leaf $l_{ij}$ to its ancestor $v$ such that the leaves in the subtree rooted at $v$ coincide with the submatrices which have $B$ in their upper left corner. As we have already observed, if $(i,j)$ is {$\ell$-aligned}, it is also $\ell'$ aligned for any $\ell' \leq \ell$, so the same leaf can have more than one backward pointer (indexed by the level number). Since the number of {$\ell$-aligned} positions is $(n/k^\ell)^2$, the total number of backward pointers is
\begin{equation}\label{eq:totbp}
\sum_{i=0}^{\lmax} (n/k^i)^2 = n^2 \cdot \sum_{i=0}^{\lmax} k^{-2i} = O(n^2).    
\end{equation}
In Appendix~\ref{app:backpointers} we show that we can compute all backward pointers in $O(n^2)$ space and time. The reason for introducing the backward pointers in $IST(M)$ is to compute the horizontal pointers in the \BTDD.  
Given a level-$\ell$ unmarked block $B$ with top left corner at position $(i,j)$, we initialise its horizontal pointers in $O(1)$ time as follows: 1) we traverse the level-$\ell$ backward pointer stored in the leaf $l_{ij}$ so we reach the ancestor $v$ defined above; 2) we get the range $[s_v,e_v]$ of leaves descending from $v$ and we find the first occurrence $O$ of $B$ in row major order by performing a $RMQ$ query on $R[s_v..e_v]$; 3) from the coordinates of the upper left corner of $O$ we retrieve the (at most) four level-$\ell$ blocks $B_1,\ldots, B_4$ overlapping $O$. The nodes representing
$B_1,\ldots,B_4$ in the \BTDD are the targets of $B$'s horizontal pointers. 

We can now state the main result of this section.

\begin{theorem}\label{theorem:2DBTbuild}
Given a matrix $M\in \Sigma^{n\times n}$, the \BTDD on $M$ can be built in $O(n^2)$ time and space. 
\end{theorem}

\begin{proof}

Initially we determine which blocks are marked and we store this information in a three-dimensional matrix of bits $N$ such that is $N[i][j][\ell]=1$ if and only if position $(i,j)$ is $\ell$-aligned and its corresponding level-$\ell$ block is marked. Note that since there are $\Theta(\log n)$ levels the matrix $N$ takes $O(n^2)$ words and can be initialized to 0 in $O(n^2)$ time. Once all marked blocks have been identified, we construct a tree $T$ which is identical to the target \BTDD but without the horizontal pointers. Finally, by adding to $T$ the horizontal pointers we get the target \BTDD. Recall that a level-$\ell$ block is marked if it is contained in a level-$\ell$ superblock which is a first occurrence. Hence, to determine the marked blocks we only need to find all superblocks which are first occurrences. With our notation a level-$\ell$ superblock is a $2 k^\ell \times 2 k^\ell$ submatrix which starts in a $\ell$-aligned position. We can find the desired superblocks with a visit of $IST(M)$ in which every time we traverse a tree arc $v' \to v$ we proceed as follows. Let $[s_v,e_v]$ denote the range of leaves in the subtree rooted at $v$, and let $l_{ij} = X[\min R[s_v..e_v]]$ denote the leaf in $v$'s subtree appearing first in row major order. For every value $\ell$ such that: 1) the size of the matrix associated to the parent node $v'$ has size strictly smaller than $2 k^\ell \times 2 k^\ell$, 2) the size of the matrix associated to $v$ is at least $2 k^\ell \times 2 k^\ell$, and 3) position $(i,j)$ is $\ell$-aligned, we have that the $2 k^\ell \times 2 k^\ell$ superblock $D$ starting at $(i,j)$ is a first occurrence, hence the four level-$\ell$ blocks contained in $D$ are marked by setting the corresponding bit in matrix $N$. It is easy to see that the values $\ell$ that satisfies conditions 1--3 above are a range of consecutive values $\ell', \ell'+1, \ldots, \ell' + h$, so as soon as a value $\ell$ fails one of the above conditions we can stop considering $l_{ij}$ and resume the visit using that value $\ell$ as a first candidate for $v$'s children.
Because of the above properties, the overall cost of finding the marked blocks and initializing~$N$ is bounded by the number of superblocks plus the number of nodes in $IST(M)$ which are both $O(n^2)$. 

Once we have initialized $N$ we can build the \BTDD starting from the root and proceeding level by level, initially without storing the horizontal pointers. After this initial construction, we traverse the resulting tree structure and every time we find an unmarked leaf we initialise its horizontal pointers using the procedure outlined above. We only need to take care of a minor detail: the procedure outlined above gives us the upper left corners of the target blocks, while in the \BTDD we need the pointers the nodes which represent those blocks. A simple solution to this problem without introducing additional storage is the following: every time a node representing a marked level-$\ell$ block starting in $(i,j)$ is added to the \BTDD, we store a pointer to that node in the leaf $l_{ij}$ overwriting the level-$\ell$ backward pointer. Such backward pointer exists by construction and we can overwrite it since we use backward pointers only to find the first occurrence of {\em unmarked} blocks. \qed
\end{proof}

We now show that we can build in linear time and space also the attractor-based \BTDD starting from a two-dimensional attractor $\GammaDD$. The procedure we propose differs from that of Theorem~\ref{theorem:2DBTbuild} in two crucial points: 1) how we find the marked blocks, 2) how we identify the target nodes of the horizontal pointers, i.e. given a unmarked block $B$ how we find one of its occurrences which crosses an attractor position. 
For the construction we augment $IST(M)$ with the extra information used for the \BTDD construction of Theorem~\ref{theorem:2DBTbuild} with exception of~$R$. In addition, we store a matrix $L$ such that $L[i][j]$ is the minimum side length of a square submatrix of $M$ with top-left corner at position $(i,j)$ which includes at least one attractor position, or $+\infty$ if such a submatrix does not exist. In addition, we store the array $\Delta[1..n^2]$ such that $\Delta[p]$ is the value $L[i][j]$ of the $p$-th leaf $l_{ij}$ in lexicographic order and we equip $\Delta$ with a data structure supporting constant time \textrm{RMQ} queries~\cite{rmqFischerJohannes}.

% We solve 2) in $O(1)$ time as in Theorem~\ref{theorem:2DBTbuild}, by just replacing the rank values stored in the leaves-aligned array $R$ with least distances from an attractor position.

\begin{theorem}
Given a matrix $M\in \Sigma^{n\times n}$ and an attractor $\GammaDD$ for $M$, the attractor based \BTDD on $M$ can be built in $O(n^2)$ time and space. 
\end{theorem}

\begin{proof}\label{theorem:GTconstruction}
As in Theorem~\ref{theorem:2DBTbuild} we determine which blocks are marked and store this information in a three-dimensional bit matrix~$N$.
Note that this task is now considerably simpler since the positions of marked blocks are determined by the attractor positions and therefore they can be determined without using the Isuffix Tree. Indeed, we mark the positions in $N$ with the following two stage procedure: first we mark the entries corresponding to the blocks which include an attractor position, then we mark the entries corresponding to the (up to eight) blocks adjacent to a block marked in the first stage.
Suppose $T$ is the target attractor-based block tree. During the first stage, for any position $p$ in the attractor we should mark in $N$ the leaf corresponding to $p$ in $T$ and all its ancestors, because they correspond to all the blocks (of different levels) containing $p$.
However, if during this ``virtual'' ascent in $T$ we reach a node which is already marked, we can avoid going any further up in the tree because also its ancestors will have been certainly already marked (by the processing of another attractor position). Formally, during the first stage, for each attractor position $p$ we scan the levels $\ell$ starting from $\ell=0$ and we compute the level-$\ell$ block $B$ including $p$ in $O(1)$ time. If $B$ has not been previously marked in $N$ we mark it by setting the corresponding bit and we proceed with the next level. Otherwise, if $B$ has been already marked, we proceed with the next attractor position restarting with $\ell=0$. In the second stage we simply scan the blocks which has been marked in the previous phase and we mark their neighbours. 
To bound the running time observe that in the first stage we mark a different block at each step, with the exception of the last step of each attractor positions. Given that $|\GammaDD|\leq n^2$ the first stage takes $O(n^2)$ time. Since during the second stage we mark $O(1)$ adjacent blocks for each block marked in the preceding phase, filling the matrix $N$ takes $O(n^2)$ overall time.

After having initialized $N$ we proceed to build the \BTDD as in Theorem~\ref{theorem:2DBTbuild}. We change the algorithm only in the very final phase when we have to determine the target nodes of the horizontal pointers. Given a level-$\ell$ unmarked block $B$, we follow a backward pointer to find its ancestor $v$ such that $[s_v,e_v]$ consists of the range of leaves whose $k^\ell \times k^\ell$ submatrix in the upper left corner has the same content as $B$.  By computing $\min (\Delta[s_v..e_v])$ we find the occurrence $O$ of $B$ closer to an attractor position; since $\GammaDD$ is an attractor it must be $\min(\Delta[s_v..e_v]) \leq k^\ell$. From the starting position of $O$ we compute the positions of the (up to four) level-$\ell$ blocks overlapping $O$ and we use them to set the horizontal pointers for $B$.

To complete the proof we only need to show that we can compute $L$ in $O(n^2)$ time. This can be done via dynamic programming filling $L$ bottom-up and right to left by using the following recurrence: $L[i][j]=1$ if $(i,j)\in \GammaDD$, otherwise it is 
$$
d = 1 +\min(L[i+1][j],L[i][j+1],L[i+1][j+1])
$$
and
$$
L[i][j] = 
\begin{cases}
    +\infty  &  \text{ if } \max(i+d,j+d)> n+1\\
   d & \text{otherwise;}
\end{cases}  
$$
where we assume $L[i][j]=+\infty$ when $(i>n) \vee (j>n)$. The case $\max(i+d,j+d)>n+1$ in the above formula covers the possibility in which there exists $p \in \GammaDD$ at distance $d$ from $(i,j)$, but the $d\times d$ square starting at $(i,j)$ is not entirely contained within $M$. \qed
\end{proof}

For a general $n$ which is not a power of $k$, we can easily adapt this construction algorithm to handle the rectangular blocks in the last row/column of $M$ (see Figure~\ref{fig:superblockR}). In this scenario, as outlined in Section~\ref{sec:2DBT},  we only need to modify the horizontal pointers computation for rectangular blocks as follows. Given a rectangular unmarked block $R$, we consider the smallest square submatrix $S$ including $R$, and we compute $R$'s horizontal pointers and offset by searching for an occurrence $S'$ of $S$ such that $S'$ contains an attractor position. We can find $S'$ in $O(1)$ time as described above for the case $n$ power of $k$, by just storing the backward pointer for the square submatrix $S$ instead of the one for the rectangular block $R$. This can be done with the procedure described in Appendix~\ref{app:backpointers} by simply considering $S$'s starting position as an aligned position for the corresponding level.

%The number of extra backward pointers is equal to the number of rectangular blocks which is, summing over the levels of tree, $O(n)$ and can be computed as described in the Appendix~\ref{app:backpointers} 

\section{Concluding Remarks}\label{sec:concluding}

We have introduced three compressibility measures for two-dimensional data that generalize some measures recently introduced for strings. We have proven some relationship between such two-dimensional compressibility measures and found that, somewhat surprisingly, there are properties that hold for strings that do not generalize in two dimensions. We have established a strong connection between our measures and the two-dimensional block tree showing that the space usage of the block tree can be bounded in terms of such measures.

Our measures are based on combinatorial properties of the square submatrices of the input. After the publication of the preliminary version of our paper~\cite{CarfagnaM23}, Romana et al.~\cite{Romana_Sciortino_Urbina_2024} proposed and analyzed 2D-measures analogous to ours but considering {\em rectangular} submatrices of any size; in the same paper they also introduced two grammar-based 2D-measures. It turns out that most of the relationship we have established for our measures are valid even in the more general context of rectangular submatrices. Although measures based on rectangular submatrices certainly have a theoretical appeal, from the algorithmic point of view square submatrices appear to be more manageable, as witnessed by the two-dimensional block tree and the Isuffix Tree data structures which are both based on square submatrices.

Our results on two dimensional measures lead to some interesting open questions, both of theoretical and practical flavour. For example:
\begin{itemize}
    \item The measure $\bDD$ defined in terms of the smallest two-dimensional bi-directional macro scheme can be asymptotically smaller than $\gammaDD$ and $\deltaDD$ (Theorem~\ref{theo:b<delta}) yet it is reachable, in the sense that we can represent any matrix $M$ in $O(\bDD(M))$ words. In one dimension computing the minimal bi-directional parsing is NP-complete but the Lempel-Ziv parsing is a practical alternative since it can be computed in linear time and its size differs from the optimal by a logarithmic factor~\cite[Section~3]{Navacmcs20.3}. It would be very interesting to find a two-dimensional analogous of the Lempel-Ziv parsing: that is, a two dimensional parsing that can be computed efficiently and whose size is provably not too far from $\bDD$.
    \item We have proven that there is a gap of $\Omega(\sqrt{n})$ between $\deltaDD$ and $\gammaDD$ for matrices with $\deltaDD = O(1)$ and that this gap becomes larger if we consider three-dimensional (cubic) structures. It would be interesting to investigate the relationship between $\deltaDD$ and $\gammaDD$ for matrices with an arbitrary measure $\deltaDD$ as it has been done for strings in~\cite{titt22}.
    \item We have proven that there is a logarithmic gap in the measure $\bDD$ if we consider the optimal parsing with or without overlapping. {Bidirectional macro schemes for strings do not allow overlapping: we conjecture that there is no need to introduce it since overlapping does not help in one-dimension, but could not prove this result}.
\end{itemize}

Our results on the construction of the two dimensional block-tree and on its size are an incentive to better study this data structure. We point out that we have considered relatively simple, and sometime redundant, marking strategies since in this paper we were only interested in bounding the asymptotic number of marked blocks. Indeed, at every level we have a certain degree of freedom in choosing the marked blocks and such freedom could be used to reduce the overall number of marked blocks and therefore the overall size of the structure. In addition, to make the construction more practical, it would be interesting to investigate the trade-offs we can obtain replacing the Isuffix Tree with the two-dimensional suffix array~\cite{Kim_Kim_Park_2003}, or some other simpler data structure, since in our construction we do not use the full power of the Isuffix Tree.

\bibliographystyle{plain}
\bibliography{main}

\appendix

\section{Backward pointers computation}\label{app:backpointers}

In this section we show how to compute the backward pointers in $O(n^2)$ time and space. Recall that a position $(i,j)$ is $\ell$-aligned if $i-1$ and $j-1$ are both multiple of $k^\ell$. This means that there is a value $\lambda_{ij}$ such that $(i,j)$ is $\ell$-aligned for $\ell = 0, 1, \ldots \lambda_{ij}$. For simplicity in the following if $(i,j)$ is $\ell$-aligned we also say that leaf $l_{ij}$ in $IST(M)$ is $\ell$-aligned. The backward pointers for leaf $l_{ij}$ are stored into an array $BP_{ij}[0..\lambda_{ij}]$ such that $BP_{ij}[\ell]$ points to the ancestor $v$ of $l_{ij}$ such that the leaves in the subtree rooted at $v$ coincide with the set of leaves which share with $l_{ij}$ the $k^\ell\times k^\ell$ submatrix in the upper left corner. As shown by~\eqref{eq:totbp} the total number of backward pointers is $O(n^2)$. 

To compute the arrays $BP_{ij}[0..\lambda_{ij}]$ we make use of a set of auxiliary integer arrays of the same size $LN_{ij}[0..\lambda_{ij}]$ defined as follows ($LN$ stands for Lexicographically Next). 
By construction, $LN_{ij}[\ell]$ only exists if the leaf $l_{ij}$ is $\ell$-aligned; $LN_{ij}[\ell]$ stores the lexicographic rank of the first $\ell$-aligned leaf following $l_{ij}$ in lexicographic order.
The arrays $LN_{ij}$'s can be initialized in $O(n^2)$ time with a left to right scan of the leaves. During the scan we maintain an auxiliary array $A$ such that $A[\ell]$ stores the last processed leaf whose position is $\ell$-aligned. When we encounter a new leaf $l_{ij}$ with lexicographic rank $r_{ij}$, for $\ell=0,\ldots,\lambda_{ij}$ we store $r_{ij}$ in the $LN$ of leaf $A[\ell]$ and we set $A[\ell] = l_{ij}$. During the initialization phase we also initialize an array $H$ such that $H[\ell]$ is the rank of the lexicographically first $\ell$-aligned leaf. Hence, if we start with $H[\ell]$ and follows the $LN_{ij}[\ell]$ values we traverse all the $\ell$-aligned leaves in lexicographic order. 

To compute the backward pointers we perform a visit of $IST(M)$ and every time we traverse a tree arc $v' \to v$ we proceed as follows. Let $[s_v,e_v]$ denote the range of leaves in the subtree rooted at $v$. For every value $\ell$ such that: 1) the size of the matrix associated to the parent node $v'$ has size strictly smaller than $k^\ell \times k^\ell$, 2) the size of the matrix associated to $v$ is at least $ k^\ell \times k^\ell$ we must set the backward pointers $BP_{ij}[\ell]$ to $v$ for all $\ell$-aligned leaves $l_{ij}$ whose lexicographic rank is between $s_v$ and $e_v$ (if any, it is possible that the range $[s_v,e_v]$ contains no $\ell$-aligned leaves). The crucial observation is that, if we consider a single level $\ell$, during the tree visit the $\ell$-aligned leaves are visited in lexicographic order therefore we can process them in constant time per $\ell$-aligned leaf using $H[\ell]$ and the arrays $LN_{ij}[\ell]$ to generate the ranks of $\ell$-aligned leaves as outlined above.  During the visit we process all levels simultaneously, but the overall running time is still bounded by the total number of backward pointers, which is $O(n^2)$, plus the overhead of processing the tree nodes. The cost of processing node $v$ is $O(n_v)$ where $n_v$ is the number of levels processed while visiting $v$. We assume that we maintain the last level processed by $v$'s ancestors, so we terminate processing $v$ as soon as we find a level $\ell$ for which there are no $\ell$-aligned leaves in $[s_v,e_v]$ (because this implies there cannot be $\ell'$-aligned leave in $[s_v,e_v]$ even for $\ell'>\ell$). Hence, among the $n_v$ processed levels, $n_v-1$ of them caused at least one backward pointer to be written so the cost of that levels is amortized by the backward pointer writing; only for the last level no backward pointer is written, but this is a constant overhead per node and the whole computation takes $O(n^2)$ time as claimed.

\section{Construction of the 2D block tree for arbitrary matrix size}\label{app:nopow}

{In} this appendix we show how to adapt the \BTDD construction algorithm of Theorem~\ref{theorem:2DBTbuild} for an $n\times n$ matrix when $n$ is not a power of $k$. As we already observed (see Figure~\ref{fig:superblockR}) in this case at each level there can be rectangular blocks along the right and bottom border of $M$.

Our construction algorithm uses the Isuffix Tree $IST(M)$ to 1) determine which blocks are marked at each level and 2) find the first occurrences of the unmarked blocks. Since the Isuffix Tree enables efficient pattern matching only for square submatrices, to handle the general case we slightly adjust the marking scheme outlined in Section~\ref{sec:2DBT} in order to operate only with square submatrices even in the presence of rectangular blocks. Notice that the adjustments involve only the blocks and superblocks on the bottom or right edge of $M$ which are the only ones that can have a rectangular shape. 
Consider any rectangular superblock $R$. Since we cannot efficiently check whether $R$ is a block-marker, we consider instead the smallest square submatrix $S$ containing $R$; for example, in Figure~\ref{fig:superblockR} $S$ is the square submatrix highlighted in gray.
We use the Isuffix Tree to check if $S$ is a first occurrence and we mark the blocks in $R$ if this is the case. It's important to note that this approach may result in a few extra blocks being marked compared to the strategy outlined in Section~\ref{sec:2DBT}, as $S$ could be a first occurrence even if $R$ is not (but not vice versa!). To bound the number of additional marked blocks we observe that the number of this special kind of block-markers (i.e. those containing a rectangular superblock) can be bounded as the type 2 block-markers in the proof of Lemma~\ref{lemma:2DBTnodes}. Hence, the results of Theorem~\ref{theorem:2DBTspace} still hold.

In order to implement the above idea, the construction algorithm in Theorem~\ref{theorem:2DBTbuild} only need to be modified as follows. During the traversal of the Isuffix Tree to determine which blocks are marked, we have also to consider the square submatrices $S$ including a rectangular superblock $R$ (e.g. the gray submatrix in Figure~\ref{fig:superblockR}). To this end, when we traverse an arc, we also consider the values $\ell$ which satisfy the conditions 1) and 2) of Theorem~\ref{theorem:2DBTbuild} and the following additional constraint: 3) the position $(i,j)$ of the descending leaf $l_{ij}$ appearing first in row major order is a possible starting position of a level-$\ell$ square $S$, because in such case we have to mark the $4$ blocks included in the corresponding rectangular superblock $R$. Since a level-$\ell$ square submatrix $S$ defined as above could start only at positions of the form $(1+\lambda k^\ell, n-2k^\ell+1)$, or $(n-2k^\ell+1, 1+\lambda k^\ell)$ or $(n-2k^\ell+1, n-2k^\ell+1)$, this last requirement holds if and only if position $(i,j)$ matches one of these three forms. We note that, since for each traversed arc there is at most one value $\ell^*$ that meets all the conditions 1--3 above and we can compute it in $O(1)$ time, the traversal still takes $O(n^2)$ overall time.

Finally, to compute the horizontal pointers of a rectangular unmarked block $R$, we consider the smallest square submatrix $S$ containing $R$ and we use the Isuffix tree to find the first occurrence $S'$ of $S$. By construction $S'$ contains a submatrix $R'$ equal to $R$ and the blocks overlapping $S'$ (and thus $R'$) are marked. Hence $R'$ can be used as target for the horizontal pointers of the unmarked node corresponding to $R$. To implement the above strategy we modify the construction so that instead of storing the backward pointers for the rectangular blocks, we store those for the smallest square submatrices containing them. These backward pointers can be computed using the procedure described in Appendix~\ref{app:backpointers}, by taking as $\ell$-aligned positions the starting positions of the above square submatrices instead of those of the rectangular blocks they contain.

\end{document}